\documentclass[12pt]{article}
\usepackage{amsfonts}
\usepackage[dvips]{graphicx}
\usepackage{latexsym,amsmath,amssymb,graphics,stmaryrd}
\usepackage{array}
\usepackage{subfigure}
\usepackage{multirow}
\usepackage[dvips]{graphicx}
\usepackage[dvips]{color}
\usepackage{pstricks}
\usepackage{pst-node}
\usepackage{dsfont}
\usepackage{amsthm}
\usepackage{graphics}
\usepackage{subfigure}
\usepackage{fancyhdr}
\usepackage[dvips]{color}
\newcommand{\col}{~,}
\newcommand{\pnt}{~.}
\newcommand{\AdS}{\text{AdS}}

\newcommand{\twob}{{\text{II}\,\text{B}}}
\newcommand{\V}[3]{V^{#1(#2)}_{#3}}
\newcommand{\Vw}[3]{V^{#1(#2)}_{\text{w},#3}}
\newcommand{\Vnw}[3]{V^{#1(#2)}_{\text{nw},#3}}

\newcommand{\D}[3]{D^{#1(#2)}_{#3}}
\newcommand{\Dw}[3]{D^{#1(#2)}_{\text{w},#3}}
\newcommand{\Dnw}[3]{D^{#1(#2)}_{\text{nw},#3}}

\newcommand{\f}[2]{f^{(#1)}_{#2}}
\newcommand{\fw}[2]{f^{(#1)}_{\text{w},#2}}

\renewcommand{\O}[2]{\mathcal{O}^{#1}_{#2}}
\newcommand{\de}{\operatorname{d}\!}
\newgray{ogray}{0.85}
\newgray{hatchgray}{1}
\newgray{sgray}{0.8}
\newcommand{\olcolor}{ogray}
\newcommand{\olfillstyle}{crosshatch*}
\newcommand{\olhatchcolor}{hatchgray}
\newcommand{\orcolor}{ogray}
\newcommand{\orfillstyle}{crosshatch*}
\newcommand{\orhatchcolor}{hatchgray}
\newcommand{\sfillstyle}{solid}

\newlength{\unit}
\newlength{\rad}
\newlength{\roff}
\newlength{\ri}
\psset{xunit=\unit,yunit=\unit,runit=\unit}
\newlength{\dlinewidth}
\newlength{\linew}
\newlength{\doublesep}
\psset{doublesep=\doublesep}
\psset{linewidth=\linew}
\newlength{\linearc}
\newlength{\xa}
\newlength{\ya}
\newlength{\xb}
\newlength{\yb}
\newlength{\xc}
\newlength{\yc}
\newlength{\xd}
\newlength{\yd}
\newcommand{\ulinsert}[3][white]{%
\setlength{\xa}{#2\unit}
\addtolength{\xa}{0\unit}
\setlength{\xb}{#2\unit}
\addtolength{\xb}{1.5\unit}
\setlength{\xc}{#2\unit}
\addtolength{\xc}{3\unit}
\setlength{\ya}{#3\unit}
\addtolength{\ya}{0.5\dlinewidth}
\setlength{\yb}{#3\unit}
\addtolength{\yb}{-0.5\dlinewidth}
\setlength{\yc}{#3\unit}
\addtolength{\yc}{-1.5\unit}
\psset{doubleline=false}
\pscustom[fillstyle=\sfillstyle,fillcolor=#1,linecolor=#1,linewidth=0pt]{%
\psline[liftpen=1,linearc=\linearc](\xc,\yb)(\xb,\yb)(\xb,\yc)
\psline[liftpen=1](\xb,\ya)(\xc,\ya)}
\pscustom[fillstyle=\olfillstyle,fillcolor=\olcolor,hatchcolor=\olhatchcolor,
linecolor=\olcolor,linewidth=\linew]{%
\psline[linearc=\linearc](\xb,\yc)(\xb,\ya)
\psline[liftpen=1,linearc=2\linearc](\xb,\ya)(\xa,\ya)(\xa,\yc)}
\psline[linearc=\linearc](\xb,\yc)(\xb,\yb)(\xc,\yb)
\psline[linearc=2\linearc](\xa,\yc)(\xa,\ya)(\xb,\ya)
\psline(\xb,\ya)(\xc,\ya)
}
\newcommand{\dlinsert}[3][white]{%
\setlength{\xa}{#2\unit}
\addtolength{\xa}{0\unit}
\setlength{\xb}{#2\unit}
\addtolength{\xb}{1.5\unit}
\setlength{\xc}{#2\unit}
\addtolength{\xc}{3\unit}
\setlength{\ya}{#3\unit}
\addtolength{\ya}{-0.5\dlinewidth}
\setlength{\yb}{#3\unit}
\addtolength{\yb}{0.5\dlinewidth}
\setlength{\yc}{#3\unit}
\addtolength{\yc}{1.5\unit}
\psset{doubleline=false}
\pscustom[fillstyle=\sfillstyle,fillcolor=#1,linecolor=#1,linewidth=0pt]{%
\psline[linearc=\linearc](\xc,\yb)(\xb,\yb)(\xb,\yc)
\psline(\xb,\ya)(\xc,\ya)}
\pscustom[fillstyle=\olfillstyle,fillcolor=\olcolor,hatchcolor=\olhatchcolor,
linecolor=\olcolor,linewidth=\linew]{%
\psline[linearc=\linearc](\xb,\yc)(\xb,\ya)
\psline[liftpen=1,linearc=2\linearc](\xb,\ya)(\xa,\ya)(\xa,\yc)}
\psline[linearc=\linearc](\xb,\yc)(\xb,\yb)(\xc,\yb)
\psline[linearc=2\linearc](\xa,\yc)(\xa,\ya)(\xb,\ya)
\psline(\xb,\ya)(\xc,\ya)
}
\newcommand{\drinsert}[3][white]{%
\setlength{\xa}{#2\unit}
\addtolength{\xa}{0\unit}
\setlength{\xb}{#2\unit}
\addtolength{\xb}{-1.5\unit}
\setlength{\xc}{#2\unit}
\addtolength{\xc}{-3\unit}
\setlength{\ya}{#3\unit}
\addtolength{\ya}{-0.5\dlinewidth}
\setlength{\yb}{#3\unit}
\addtolength{\yb}{0.5\dlinewidth}
\setlength{\yc}{#3\unit}
\addtolength{\yc}{1.5\unit}
\psset{doubleline=false}
\pscustom[fillstyle=\sfillstyle,fillcolor=#1,linecolor=#1,linewidth=0pt]{%
\psline[linearc=\linearc](\xc,\yb)(\xb,\yb)(\xb,\yc)
\psline(\xb,\ya)(\xc,\ya)}
\pscustom[fillstyle=\orfillstyle,fillcolor=\orcolor,hatchcolor=\orhatchcolor,
linecolor=\orcolor,linewidth=\linew]{%
\psline[linearc=\linearc](\xb,\yc)(\xb,\ya)
\psline[liftpen=1,linearc=2\linearc](\xb,\ya)(\xa,\ya)(\xa,\yc)}
\psline[linearc=\linearc](\xb,\yc)(\xb,\yb)(\xc,\yb)
\psline[linearc=2\linearc](\xa,\yc)(\xa,\ya)(\xb,\ya)
\psline(\xb,\ya)(\xc,\ya)
}
\newcommand{\urinsert}[3][white]{%
\setlength{\xa}{#2\unit}
\addtolength{\xa}{0\unit}
\setlength{\xb}{#2\unit}
\addtolength{\xb}{-1.5\unit}
\setlength{\xc}{#2\unit}
\addtolength{\xc}{-3\unit}
\setlength{\ya}{#3\unit}
\addtolength{\ya}{0.5\dlinewidth}
\setlength{\yb}{#3\unit}
\addtolength{\yb}{-0.5\dlinewidth}
\setlength{\yc}{#3\unit}
\addtolength{\yc}{-1.5\unit}
\psset{doubleline=false}
\pscustom[fillstyle=\sfillstyle,fillcolor=#1,linecolor=#1,linewidth=0pt]{%
\psline[linearc=\linearc](\xc,\yb)(\xb,\yb)(\xb,\yc)
\psline[liftpen=1](\xb,\ya)(\xc,\ya)}
\pscustom[fillstyle=\orfillstyle,fillcolor=\orcolor,hatchcolor=\orhatchcolor,
linecolor=\orcolor,linewidth=\linew]{%
\psline[linearc=\linearc](\xb,\yc)(\xb,\ya)
\psline[liftpen=1,linearc=2\linearc](\xb,\ya)(\xa,\ya)(\xa,\yc)}
\psline[linearc=\linearc](\xb,\yc)(\xb,\yb)(\xc,\yb)
\psline[linearc=2\linearc](\xa,\yc)(\xa,\ya)(\xb,\ya)
\psline(\xb,\ya)(\xc,\ya)
}
\newcommand{\olvertex}[3][white]{%
\setlength{\xa}{#2\unit}
\addtolength{\xa}{0\unit}
\setlength{\xb}{#2\unit}
\addtolength{\xb}{1.5\unit}
\setlength{\xc}{#2\unit}
\addtolength{\xc}{3\unit}
\setlength{\ya}{#3\unit}
\addtolength{\ya}{1.5\unit}
\setlength{\yb}{#3\unit}
\addtolength{\yb}{0.5\dlinewidth}
\setlength{\yc}{#3\unit}
\addtolength{\yc}{-0.5\dlinewidth}
\setlength{\yd}{#3\unit}
\addtolength{\yd}{-1.5\unit}
\psset{doubleline=false}
\pscustom[fillstyle=\sfillstyle,fillcolor=#1,linecolor=#1,linewidth=0pt]{%
\psline[linearc=\linearc](\xc,\yb)(\xb,\yb)(\xb,\ya)
\psline[liftpen=1,linearc=\linearc](\xb,\yd)(\xb,\yc)(\xc,\yc)}
\pscustom[fillstyle=\olfillstyle,fillcolor=\olcolor,hatchcolor=\olhatchcolor,
linecolor=\olcolor,linewidth=0pt]{%
\psline[liftpen=0](\xa,\yd)(\xb,\yd)
\psline[liftpen=0](\xb,\ya)(\xa,\ya)
}
\psline[linecolor=\olcolor,linewidth=\linew](\xb,\ya)(\xb,\yd)
\psline[linearc=\linearc]{-C}(\xc,\yb)(\xb,\yb)(\xb,\ya)
\psline[liftpen=1,linearc=\linearc](\xb,\yd)(\xb,\yc)(\xc,\yc)
\psline{C-}(\xa,\ya)(\xa,\yd)
}
\newcommand{\orvertex}[3][white]{%
\setlength{\xa}{#2\unit}
\addtolength{\xa}{0\unit}
\setlength{\xb}{#2\unit}
\addtolength{\xb}{-1.5\unit}
\setlength{\xc}{#2\unit}
\addtolength{\xc}{-3\unit}
\setlength{\ya}{#3\unit}
\addtolength{\ya}{1.5\unit}
\setlength{\yb}{#3\unit}
\addtolength{\yb}{0.5\dlinewidth}
\setlength{\yc}{#3\unit}
\addtolength{\yc}{-0.5\dlinewidth}
\setlength{\yd}{#3\unit}
\addtolength{\yd}{-1.5\unit}
\psset{doubleline=false}
\pscustom[fillstyle=\sfillstyle,fillcolor=#1,linecolor=#1,linewidth=0pt]{%
\psline[linearc=\linearc](\xc,\yb)(\xb,\yb)(\xb,\ya)
\psline[liftpen=1,linearc=\linearc](\xb,\yd)(\xb,\yc)(\xc,\yc)}
\pscustom[fillstyle=\orfillstyle,fillcolor=\orcolor,hatchcolor=\orhatchcolor,
linecolor=\orcolor,linewidth=0pt]{%
\psline[liftpen=0](\xb,\ya)(\xa,\ya)
\psline[liftpen=0](\xa,\yd)(\xb,\yd)
\psline[liftpen=0](\xb,\ya)(\xa,\ya)
}
\psline[linecolor=\orcolor,linewidth=\linew](\xb,\ya)(\xb,\yd)
\psline[linearc=\linearc]{-C}(\xc,\yb)(\xb,\yb)(\xb,\ya)
\psline[liftpen=1,linearc=\linearc](\xb,\yd)(\xb,\yc)(\xc,\yc)
\psline{C-}(\xa,\ya)(\xa,\yd)
}
\newcommand{\oldots}[3][white]{%
\setlength{\xa}{#2\unit}
\addtolength{\xa}{0\unit}
\setlength{\xb}{#2\unit}
\addtolength{\xb}{1.5\unit}
\setlength{\xc}{#2\unit}
\addtolength{\xc}{3\unit}
\setlength{\ya}{#3\unit}
\addtolength{\ya}{1.5\unit}
\setlength{\yb}{#3\unit}
\addtolength{\yb}{0.5\dlinewidth}
\setlength{\yc}{#3\unit}
\addtolength{\yc}{-0.5\dlinewidth}
\setlength{\yd}{#3\unit}
\addtolength{\yd}{-1.5\unit}
\psset{doubleline=false}
\pscustom[fillstyle=\sfillstyle,fillcolor=#1,linecolor=#1,linewidth=0pt]{%
\psline[liftpen=1](\xb,\yd)(\xb,\ya)}
\pscustom[fillstyle=\olfillstyle,fillcolor=\olcolor,hatchcolor=\olhatchcolor,
linecolor=\olcolor,linewidth=0pt]{%
\psline[liftpen=0](\xa,\yd)(\xb,\yd)
\psline[liftpen=0](\xb,\ya)(\xa,\ya)
}
\psline[linecolor=\olcolor,linewidth=\linew](\xb,\ya)(\xb,\yd)
\psline[liftpen=1,linestyle=dotted](\xb,\yd)(\xb,\ya)
\psline{C-}(\xa,\ya)(\xa,\yd)
}
\newcommand{\ordots}[3][white]{%
\setlength{\xa}{#2\unit}
\addtolength{\xa}{0\unit}
\setlength{\xb}{#2\unit}
\addtolength{\xb}{-1.5\unit}
\setlength{\xc}{#2\unit}
\addtolength{\xc}{-3\unit}
\setlength{\ya}{#3\unit}
\addtolength{\ya}{1.5\unit}
\setlength{\yb}{#3\unit}
\addtolength{\yb}{0.5\dlinewidth}
\setlength{\yc}{#3\unit}
\addtolength{\yc}{-0.5\dlinewidth}
\setlength{\yd}{#3\unit}
\addtolength{\yd}{-1.5\unit}
\psset{doubleline=false}
\pscustom[fillstyle=\sfillstyle,fillcolor=#1,linecolor=#1,linewidth=0pt]{%
\psline[liftpen=1](\xb,\yd)(\xb,\ya)}
\pscustom[fillstyle=\orfillstyle,fillcolor=\orcolor,hatchcolor=\orhatchcolor,
linecolor=\orcolor,linewidth=0pt]{%
\psline[liftpen=0](\xb,\ya)(\xa,\ya)
\psline[liftpen=0](\xa,\yd)(\xb,\yd)
\psline[liftpen=0](\xb,\ya)(\xa,\ya)
}
\psline[linecolor=\orcolor,linewidth=\linew](\xb,\ya)(\xb,\yd)
\psline[liftpen=1,linestyle=dotted](\xb,\yd)(\xb,\ya)
\psline{C-}(\xa,\ya)(\xa,\yd)
}
\newcommand{\threevertex}[4][white]{%
\setlength{\xa}{0\unit}
\addtolength{\xa}{-0.5\dlinewidth}
\setlength{\xb}{0\unit}
\addtolength{\xb}{0.5\dlinewidth}
\setlength{\xc}{0\unit}
\addtolength{\xc}{1\unit}
\setlength{\ya}{0\unit}
\addtolength{\ya}{1\unit}
\setlength{\yb}{0\unit}
\addtolength{\yb}{0.5\dlinewidth}
\setlength{\yc}{0\unit}
\addtolength{\yc}{-0.5\dlinewidth}
\setlength{\yd}{0\unit}
\addtolength{\yd}{-1\unit}
\psset{doubleline=false}
\rput{0}(#2\unit,#3\unit){%
\pscustom[fillstyle=\sfillstyle,fillcolor=#1,linecolor=#1,linewidth=0pt]{%
\rotate{#4}
\psline[liftpen=1,linearc=\linearc](\xb,\ya)(\xb,\yb)(\xc,\yb)
\psline[liftpen=1,linearc=\linearc](\xc,\yc)(\xb,\yc)(\xb,\yd)
\psline[liftpen=1](\xa,\yd)(\xa,\ya)}
\pscustom{%
\rotate{#4}
\psline[liftpen=2,linearc=\linearc](\xb,\ya)(\xb,\yb)(\xc,\yb)
\psline[liftpen=2,linearc=\linearc](\xc,\yc)(\xb,\yc)(\xb,\yd)
\psline[liftpen=2](\xa,\yd)(\xa,\ya)
}
}
}
\newcommand{\fourvertex}[4][white]{%
\setlength{\xa}{0\unit}
\addtolength{\xa}{-1\unit}
\setlength{\xb}{0\unit}
\addtolength{\xb}{-0.5\dlinewidth}
\setlength{\xc}{0\unit}
\addtolength{\xc}{0.5\dlinewidth}
\setlength{\xd}{0\unit}
\addtolength{\xd}{1\unit}
\setlength{\ya}{0\unit}
\addtolength{\ya}{-1\unit}
\setlength{\yb}{0\unit}
\addtolength{\yb}{-0.5\dlinewidth}
\setlength{\yc}{0\unit}
\addtolength{\yc}{0.5\dlinewidth}
\setlength{\yd}{0\unit}
\addtolength{\yd}{1\unit}
\psset{doubleline=false}
\rput{0}(#2\unit,#3\unit){%
\pscustom[fillstyle=\sfillstyle,fillcolor=#1,linecolor=#1,linewidth=0pt]{%
\rotate{#4}
\psline[liftpen=1,linearc=\linearc](\xc,\ya)(\xc,\yb)(\xd,\yb)
\psline[liftpen=1,linearc=\linearc](\xd,\yc)(\xc,\yc)(\xc,\yd)
\psline[liftpen=1,linearc=\linearc](\xb,\yd)(\xb,\yc)(\xa,\yc)
\psline[liftpen=1,linearc=\linearc](\xa,\yb)(\xb,\yb)(\xb,\ya)}
\pscustom{%
\rotate{#4}
\psline[liftpen=1,linearc=\linearc](\xc,\ya)(\xc,\yb)(\xd,\yb)
\psline[liftpen=2,linearc=\linearc](\xd,\yc)(\xc,\yc)(\xc,\yd)
\psline[liftpen=2,linearc=\linearc](\xb,\yd)(\xb,\yc)(\xa,\yc)
\psline[liftpen=2,linearc=\linearc](\xa,\yb)(\xb,\yb)(\xb,\ya)
}
}
}
\newcommand{\uoutex}[3][white]{%
\setlength{\xb}{#2\unit}
\addtolength{\xb}{-1.5\unit}
\setlength{\xc}{#2\unit}
\setlength{\ya}{#3\unit}
\addtolength{\ya}{0.5\dlinewidth}
\setlength{\yb}{#3\unit}
\addtolength{\yb}{-0.5\dlinewidth}
\setlength{\yc}{#3\unit}
\addtolength{\yc}{-1.5\unit}
\psset{doubleline=false}
\pscustom[fillstyle=\sfillstyle,fillcolor=#1,linecolor=#1]{%
\psline[liftpen=1,linearc=\linearc](\xc,\yb)(\xb,\yb)(\xb,\yc)
\psline[liftpen=1](\xb,\ya)(\xc,\ya)}
\psline[linearc=\linearc](\xb,\yc)(\xb,\yb)(\xc,\yb)
\psline(\xb,\ya)(\xc,\ya)
}
\newcommand{\doutex}[3][white]{%
\setlength{\xb}{#2\unit}
\addtolength{\xb}{-1.5\unit}
\setlength{\xc}{#2\unit}
\setlength{\ya}{#3\unit}
\addtolength{\ya}{-0.5\dlinewidth}
\setlength{\yb}{#3\unit}
\addtolength{\yb}{0.5\dlinewidth}
\setlength{\yc}{#3\unit}
\addtolength{\yc}{1.5\unit}
\psset{doubleline=false}
\pscustom[fillstyle=\sfillstyle,fillcolor=#1,linecolor=#1]{%
\psline[linearc=\linearc](\xc,\yb)(\xb,\yb)(\xb,\yc)
\psline(\xb,\ya)(\xc,\ya)}
\psline[linearc=\linearc](\xb,\yc)(\xb,\yb)(\xc,\yb)
\psline(\xb,\ya)(\xc,\ya)
}
\newcommand{\dinex}[3][white]{%
\setlength{\xb}{#2\unit}
\addtolength{\xb}{1.5\unit}
\setlength{\xc}{#2\unit}
\setlength{\ya}{#3\unit}
\addtolength{\ya}{-0.5\dlinewidth}
\setlength{\yb}{#3\unit}
\addtolength{\yb}{0.5\dlinewidth}
\setlength{\yc}{#3\unit}
\addtolength{\yc}{1.5\unit}
\psset{doubleline=false}
\pscustom[fillstyle=\sfillstyle,fillcolor=#1,linecolor=#1]{%
\psline[linearc=\linearc](\xc,\yb)(\xb,\yb)(\xb,\yc)
\psline(\xb,\ya)(\xc,\ya)}
\psline[linearc=\linearc](\xb,\yc)(\xb,\yb)(\xc,\yb)
\psline(\xb,\ya)(\xc,\ya)
}
\newcommand{\uinex}[3][white]{%
\setlength{\xb}{#2\unit}
\addtolength{\xb}{1.5\unit}
\setlength{\xc}{#2\unit}
\setlength{\ya}{#3\unit}
\addtolength{\ya}{0.5\dlinewidth}
\setlength{\yb}{#3\unit}
\addtolength{\yb}{-0.5\dlinewidth}
\setlength{\yc}{#3\unit}
\addtolength{\yc}{-1.5\unit}
\psset{doubleline=false}
\pscustom[fillstyle=\sfillstyle,fillcolor=#1,linecolor=#1]{%
\psline[linearc=\linearc](\xc,\yb)(\xb,\yb)(\xb,\yc)
\psline[liftpen=1](\xb,\ya)(\xc,\ya)}
\psline[linearc=\linearc](\xb,\yc)(\xb,\yb)(\xc,\yb)
\psline(\xb,\ya)(\xc,\ya)
}
\newcommand{\ioutex}[3][white]{%
\setlength{\xb}{#2\unit}
\addtolength{\xb}{-1.5\unit}
\setlength{\xc}{#2\unit}
\setlength{\ya}{#3\unit}
\addtolength{\ya}{1.5\unit}
\setlength{\yb}{#3\unit}
\addtolength{\yb}{0.5\dlinewidth}
\setlength{\yc}{#3\unit}
\addtolength{\yc}{-0.5\dlinewidth}
\setlength{\yd}{#3\unit}
\addtolength{\yd}{-1.5\unit}
\psset{doubleline=false}
\pscustom[fillstyle=\sfillstyle,fillcolor=#1,linecolor=#1]{%
\psline[linearc=\linearc](\xc,\yb)(\xb,\yb)(\xb,\ya)
\psline[liftpen=1,linearc=\linearc](\xb,\yd)(\xb,\yc)(\xc,\yc)}
\psline[linecolor=#1,linewidth=\linew](\xb,\ya)(\xb,\yd)
\psline[linearc=\linearc]{-C}(\xc,\yb)(\xb,\yb)(\xb,\ya)
\psline[liftpen=1,linearc=\linearc](\xb,\yd)(\xb,\yc)(\xc,\yc)
}
\newcommand{\iinex}[3][white]{%
\setlength{\xb}{#2\unit}
\addtolength{\xb}{1.5\unit}
\setlength{\xc}{#2\unit}
\setlength{\ya}{#3\unit}
\addtolength{\ya}{1.5\unit}
\setlength{\yb}{#3\unit}
\addtolength{\yb}{0.5\dlinewidth}
\setlength{\yc}{#3\unit}
\addtolength{\yc}{-0.5\dlinewidth}
\setlength{\yd}{#3\unit}
\addtolength{\yd}{-1.5\unit}
\psset{doubleline=false}
\pscustom[fillstyle=\sfillstyle,fillcolor=#1,linecolor=#1,linewidth=0pt]{%
\psline[linearc=\linearc](\xc,\yb)(\xb,\yb)(\xb,\ya)
\psline[liftpen=1,linearc=\linearc](\xb,\yd)(\xb,\yc)(\xc,\yc)}
\psline[linecolor=#1,linewidth=\linew](\xb,\ya)(\xb,\yd)
\psline[linearc=\linearc]{-C}(\xc,\yb)(\xb,\yb)(\xb,\ya)
\psline[liftpen=1,linearc=\linearc](\xb,\yd)(\xb,\yc)(\xc,\yc)
}
\newcommand{\sixdotsvertex}[4][white]{%
\SpecialCoor
\setlength{\roff}{1.30656\dlinewidth}
\setlength{\rad}{3\unit}
\setlength{\ri}{\rad}
\addtolength{\ri}{-1.30656\dlinewidth}
\pscustom[fillstyle=\sfillstyle,fillcolor=#1]{%
\gsave
\translate(#2,#3)
\rotate{#4}
\translate(\roff;22.5)
\psline[liftpen=1,linearc=\linearc](\ri;0)(0;0)(\ri;45)
\translate(\roff;202.5)
\translate(\roff;67.5)
\psline[liftpen=1,linearc=\linearc](\ri;45)(0;0)(\ri;90)
\translate(\roff;247.5)
\translate(\roff;112.5)
\psline[liftpen=1,linearc=\linearc](\ri;90)(0;0)(\ri;135)
\translate(\roff;292.5)
\translate(\roff;157.5)
\psline[liftpen=1,linearc=\linearc](\ri;135)(0;0)(\ri;180)
\translate(\roff;337.5)
\translate(\roff;202.5)
\psline[liftpen=1,linearc=\linearc](\ri;180)(0;0)(\ri;225)
\translate(\roff;22.5)
\translate(\roff;247.5)
\psline[liftpen=1,linearc=\linearc](\ri;225)(0;0)(\ri;270)
\translate(\roff;67.5)
\translate(\roff;292.5)
\psline[liftpen=1,linearc=\linearc](\ri;270)(0;0)(\ri;315)
\translate(\roff;112.5)
\translate(\roff;337.5)
\psline[liftpen=1,linearc=\linearc](\ri;315)(0;0)(\ri;0)
\translate(\roff;157.5)
\translate(-#2,-#3)
\fill
\grestore}
\SpecialCoor
\pscustom{%
\translate(#2,#3)
\rotate{#4}
\translate(\roff;67.5)
\psline[liftpen=2,linearc=\linearc](\ri;45)(0;0)(\ri;90)
\translate(\roff;247.5)
\translate(\roff;112.5)
\psline[liftpen=2,linearc=\linearc](\ri;90)(0;0)(\ri;135)
\translate(\roff;292.5)
\translate(\roff;157.5)
\psline[liftpen=2,linearc=\linearc](\ri;135)(0;0)(\ri;180)
\translate(\roff;337.5)
\translate(\roff;292.5)
\psline[liftpen=2,linearc=\linearc](\ri;270)(0;0)(\ri;315)
\translate(\roff;112.5)
\translate(-#2,-#3)
}
\pscustom[linestyle=dotted]{%
\addtolength{\ri}{0.707106\dlinewidth}
\setlength{\roff}{0.52\roff}
\translate(#2,#3)
\rotate{#4}
\translate(\roff;0)
\psline[liftpen=2,linearc=2\linearc](\ri;315)(0;0)(\ri;45)
\translate(\roff;180)
\translate(\roff;225)
\psline[liftpen=2,linearc=2\linearc](\ri;180)(0;0)(\ri;270)
\translate(\roff;45)
\addtolength{\ri}{-0.707106\dlinewidth}
\translate(-#2,-#3)
}
}

\newlength{\armlen}
\newcounter{nnodenum}%[equation]
\newcounter{mnodenum}%[equation]
\newcommand{\cpair}[3][0.15cm]{%
\settodepth{\armlen}{$#2$}%
\addtolength{\armlen}{#1}%
\addtolength{\armlen}{1.5pt}%
\rule[-\armlen]{0pt}{\armlen}%
\rnode{n}{#2}\,%
\settodepth{\armlen}{$#3$}%
\addtolength{\armlen}{#1}%
\addtolength{\armlen}{1.5pt}%
\rule[-\armlen]{0pt}{\armlen}%
\rnode{m}{#3}%
\ncbar[linewidth=0.5pt,nodesep=1pt,angle=-90,arm=#1]{n}{m}
}
\newcommand{\unitmatrix}{\mathds{1}}
\newcommand{\T}[3]{(T^{#1})^{#2}_{\phantom{#2}#3}}
\newcommand{\comm}[2]{\left[#1\smash[b]{\mathbin{,}}#2\right]}

\DeclareMathOperator{\tr}{tr}
\numberwithin{equation}{section}
\begin{document}
\begin{titlepage}
\begin{flushright}
MPP-2005-41\\ 
HU-EP-05/21\\
\end{flushright}
\mbox{ }  \hfill hep-th/0505071
\vspace{5ex}
\Large
\begin {center}     
{\bf Wrapping interactions and the genus expansion of the $2$-point function 
  of composite operators
}
\end {center}
\large
\vspace{1ex}
\begin{center}
Christoph Sieg\footnote{\label{mpi} csieg@mppmu.mpg.de} 
and Alessandro Torrielli\footnote{\label{hu} torriell@physik.hu-berlin.de}
\end{center}
\begin{center}
\ref{mpi})
Max-Planck-Institut f\"ur Physik\\
F\"ohringer Ring 6,  D-80805 M\"unchen\\[2mm]

\ref{hu})
Humboldt--Universit\"at zu Berlin, Institut f\"ur Physik\\
Newtonstra\ss e 15, D-12489 Berlin\\[2mm]  
\end{center}
\vspace{4ex}
\rm
\begin{center}
{\bf Abstract}
\end{center} 
\normalsize 
We perform a systematic analysis of wrapping interactions
for a general class of theories with color degrees of freedom, 
including $\mathcal{N}=4$ SYM.
Wrapping interactions arise in the genus expansion of the $2$-point function
of composite operators as finite size effects that start to appear at a
certain order in the coupling constant at which the range of the 
interaction is equal to the length of the operators.
We analyze in detail the relevant genus expansions, and introduce a
strategy to single out the wrapping contributions, based
on adding spectator fields. We use a toy model to 
demonstrate our procedure, performing all computations explicitly.
Although completely general, our treatment should be particularly
useful for applications to the recent problem of wrapping
contributions in some checks of the $\AdS/\text{CFT}$ correspondence.
\vfill
\end{titlepage} 
\tableofcontents
%\include{genwrap}
%\include{samplewrap}
%\include{conclusions}
%\include{appendix}
%\include{misc}
% 03.05.05 
\section{Introduction}
The $\AdS/\text{CFT}$ correspondence \cite{Maldacena:1998re}
claims that type $\twob$ string theory
on an $\AdS_5\times\text{S}^5$ background with Ramond-Ramond (RR) flux is
dual to $\mathcal{N}=4$ super Yang-Mills (SYM) theory with gauge group $SU(N)$
which is invariant under the superconformal symmetry group $SU(2,2|4)$. 
The common curvature radius $R$ of $\AdS_5$ and of $\text{S}^5$ in units of 
the string length $\sqrt{\alpha'}$ is related to the 't Hooft coupling
$\lambda$ via   
\begin{equation}
\frac{R^2}{\alpha'}=\sqrt{\lambda}\col\qquad\lambda=g^2N\col\qquad
g^2=4\pi g_\text{s}\col
\end{equation} 
where $g$ and $g_\text{s}$ are the SYM and string coupling constants,
respectively.
One important motivation for this conjecture is that the isometry group 
$SO(2,4)\times SO(6)$ of $\AdS_5\times\text{S}^5$ can be identified 
with the bosonic subgroup $SO(2,4)\times SU(4)$ of the superconformal group.
A general proof of the conjecture is still out of reach, mainly due to the
fact that a quantization of string theories in backgrounds with RR flux 
is difficult. Furthermore, the duality relates the strong coupling regime of 
one theory to the weak coupling regime of the other theory, preventing 
one from directly using perturbation theory to compute results in both 
theories that can be compared in a common regime of the parameters. 

However, it is possible to probe the correspondence in 
different limits. In one of these limits one considers classical solutions
on the string side \cite{Gubser:2002tv,Frolov:2002av} (see also the review 
\cite{Tseytlin:2003ii} and references therein).
In this case quantitative results can be computed
in the string theory as well as in its dual gauge theory.
In the regime of large quantum numbers of the string (here collectively 
denoted by the angular momentum $J$ of a rotating string solution) 
and in the large tension limit ($\sqrt{\lambda}\gg1$) the classical
energy has a regular expansion
in the modified 't Hooft coupling
$\lambda'=\frac{\lambda}{J^2}$ which is held fixed \cite{Frolov:2004bh}. 
The analysis can be extended by 
including quantum fluctuations around the classical string solution 
\cite{Frolov:2002av,Frolov:2004bh}.  
The classical string sector shows integrability, e.g. rotating 
(rigid) string solutions have been shown to be described by the classical 
integrable Neumann system \cite{Arutyunov:2003uj}.
See also \cite{Arutyunov:2003za} for a generalization and 
\cite{Alday:2005gi,Swanson:2004mk,Swanson:2004qa,Kazakov:2004qf,Arutyunov:2004yx}
for related work on integrability in this context. 

The $\AdS/\text{CFT}$ correspondence predicts a matching of the classical 
energy of closed strings with the eigenvalues of the anomalous dimension 
matrix that describes the mixing of composite operators of
$\mathcal{N}=4$ SYM under renormalization. These operators contain a 
single trace over the gauge group and can be regarded as discretized 
closed strings. In particular, the limit of
interest requires that the operators contain a huge number of fields, implying
that their mixing matrix, that has to be diagonalized, contains 
a huge number of entries. The mixing matrix itself can be obtained from the 
$2$-point functions of the composite operators. Alternatively, the
mixing problem can be recast in 
terms of an eigenvalue problem for the dilatation operator of $\mathcal{N}=4$
SYM \cite{Beisert:2003tq} (see also the reviews 
\cite{Plefka:2003nb,Beisert:2004ry} and references therein). 
Furthermore, in the planar limit, the mixing problem has been 
reformulated in terms of integrable spin chains \cite{Minahan:2002ve}. 
There, the composite single-trace operators are
regarded as cyclic spin chains. Each fundamental field `flavor' within the
trace is interpreted as a spin projection eigenvalue at the corresponding site
of the chain. The dilatation operator itself becomes the Hamiltonian of the
chain. Formulated in these terms, the 
Bethe ansatz \cite{Bethe:1931hc} (see \cite{Faddeev:1996iy} for a review) 
provides a tool for finding the energy eigenvalues 
for the spin chain states and thus the eigenvalues of the anomalous dimension
matrix.  

Remarkable agreement of 
the classical string energies with the anomalous dimensions is found
up to two loops.
However, a mismatch was observed at three loops \cite{Beisert:2003ea}. 
Another mismatch has been observed in the BMN limit \cite{Berenstein:2002jq}
of the $\AdS/\text{CFT}$ correspondence. 
This limit corresponds to an expansion around a    
pointlike classical string in the center of $\AdS_5$ that moves along a great
circle in $S^5$. In this case, moving with the pointlike string with the
velocity of light, the $\AdS_5\times\text{S}^5$ background of the 
$\AdS/\text{CFT}$ correspondence is transformed via the Penrose-G\"uven limit 
\cite{Penrose:1965,Gueven:2000ru} to the $10$-dimensional plane wave
background \cite{Blau:2001ne,Blau:2002mw}. The first  
order quantum fluctuations around the pointlike solutions are then described
by strings in this plane wave background \cite{Frolov:2002av}.  
The corrections to the energy caused by higher  
order quantum fluctuations, i.e. curvature corrections to the plane wave
background \cite{Callan:2003xr}, show a mismatch with the
corresponding anomalous dimensions starting at three loops
\cite{Callan:2004uv,Callan:2004ev,McLoughlin:2004dh}. 
There have been first indications for a mismatch between the quantum 
fluctuations at one loop order of extended string solutions, rotating
in $\text{S}^5$\cite{Frolov:2004bh} and in both, $\AdS_5$ and $\text{S}^5$
\cite{Park:2005ji}, with the respective gauge theory results
\cite{Lubcke:2004dg,Kazakov:2004nh}. However, agreement at this order was 
found later by  including an overlooked contribution to the Bethe ansatz 
\cite{Beisert:2005mq}.\footnote{We would like to thank I.\ Y.\ Park for 
bringing this fact to our attention.}  

A possible explanation for the mismatch, that does not insist on the
unsatisfactory possibility of a breakdown of
the $\AdS/\text{CFT}$ correspondence and of integrability in the planar limit, 
might be a non-commutativity of the limits taken on both sides
\cite{Serban:2004jf,Beisert:2004hm}.
On the string side, $J\to\infty$ is taken first, and then the
expression for the energy is expanded in powers of $\lambda'$.
However, on the gauge side the anomalous dimensions are computed as a 
perturbation expansion in $\lambda$ that is valid for sufficiently large but
finite $J$, and then the limit $J\to\infty$ is taken, keeping $\lambda'$
fixed. Thereby, one has ignored finite size effects, that become relevant 
whenever the spin chain length is less than the range of the interaction.
If this is the case, there are contributions in which the interaction
wraps around the state \cite{Serban:2004jf,Beisert:2004hm}. 
They are denoted as wrapping interactions. First quantitative results
including these interactions have been obtained in the context of 
a computer-based study of the plane wave matrix model at four loop
order \cite{Fischbacher:2004iu}. 

The purpose of this paper is to present a systematic and general study of
wrapping interactions, not relying on the issues of
a concrete theory and on a given perturbative order in the coupling constant. 
Instead, it contains statements valid in a class of 
theories which includes $\mathcal{N}=4$ SYM. 
Our general results are then checked at low and fixed loop order
with the help of a simple example, that is then refined in subsequent steps.
In this way we hope to have provided an initial step to 
understand the role of the wrapping interactions in the concrete challenge 
of explaining the above mentioned mismatch between semiclassical
string energies and anomalous dimensions.\\
The paper is organized as follows:

In Section \ref{2ptcop} we introduce and discuss the $2$-point function of 
composite operators, being generated by connecting the legs of the operators
with a Green function of the theory. 
We then work out the differences of the genus expansion of such a Green 
function and of the obtained $2$-point function.

In Section \ref{wrapdiag} we define and classify wrapping interactions 
as particular contributions to the Green functions that contribute at lower
genus to the $2$-point function.
We work out some general statements for wrapping interactions by adding
spectator fields to the composite operators.
A special case of particular interest in the context of the 
$\AdS/\text{CFT}$ correspondence is the case of planar contributions
to the $2$-point function. There, we find that the planar wrapping diagrams 
originate from genus one contribution to the Green functions.
We identify a unique structure for the spectator fields of planar wrapping 
diagrams.

In Section \ref{planarcontrib} we present the expressions for the effective
vertices and show their spectator structures.

In Section \ref{toymodel} we apply our results to a toy model, given
by scalar $\phi^4$ theory with a massless colored field of a single flavor. 
The toy model is then extended to multiple 
flavors and interactions with different flavor fluxes. At the end we arrive
at the $4$-scalar interaction of $\mathcal{N}=4$ SYM.

In Section \ref{candpsi} we discuss a possibility to choose 
the spectator fields in such a way that wrapping interactions can be 
automatically projected out in computer-based symbolic computations.

In Appendix \ref{countingrules} we review and summarize some counting rules
for Feynman diagrams used in the main text. We then apply them to 
extract some issues of (planar) wrapping diagrams.

In Appendix \ref{algebrarules} we
collect some useful formulae for $SU(N)$ and $U(N)$ which enter our
calculations.   

\section{The $2$-point function of composite operators}
\label{2ptcop}
\subsection{Building blocks}
Before we can define wrapping interactions we should fix our notations 
and conventions when we deal with the $2$-point functions of composite
operators.

By $\O{e}{R}$, $e=1,2$ we denote two local operators that consist of 
$r=1,\dots,R$ elementary fields $\phi_{n_r}$ where each $n_r$ denotes one
field `flavor' of the theory. Of particular interest are single-trace 
operators of the form
\begin{equation}\label{codef}
\O{}{R}=\frac{1}{\sqrt{N}^{R-2}}\tr(\phi_{n_1}\dots\phi_{n_R})\col
\end{equation}
where the trace runs over the gauge group, i.e. the elementary fields
$\phi_{n_r}$ are decomposed as $\phi_{n_r}=\phi^a_{n_r}T^a$, 
and $T^a$ are $N\times N$ representation
matrices of the Lie algebra of the gauge group. Later we will refer to the 
chosen normalization factor.

\begin{figure}
\begin{center}%
\setlength{\ya}{9\unit}%
\addtolength{\ya}{0.5\dlinewidth}%
\setlength{\yb}{0\unit}%
\addtolength{\yb}{-0.5\dlinewidth}%
\subfigure[]{\label{corep}%
\begin{pspicture}(-4,-2)(7,11)%\showgrid
%\psframe(-4,-2)(7,11)
\psframe[fillstyle=\olfillstyle,fillcolor=\olcolor,hatchcolor=\olhatchcolor,
%linecolor=\olcolor,
linewidth=\linew](0,\ya)(2,\yb)\rput[r](-1,4.5){$\textstyle\O{}{R}$}
\psline(2,9)(3,9)
\psline(2,6)(3,6)
%\psline(2,3)(3,3)
\psline(2,0)(3,0)
\psline[linestyle=dotted](3,2)(3,4)
%\psline[linestyle=dotted](4,2)(4,4)
\rput[l](3.2,0){$\scriptstyle 1$}
\rput[l](3.2,6){$\scriptstyle R-1$}
\rput[l](3.2,9){$\scriptstyle R$}
\end{pspicture}}%
\qquad\qquad
\subfigure[]{\label{Vrep}%
\begin{pspicture}(-2,-2)(14,11)%\showgrid
%\psframe(-2,-2)(14,11)
\psframe(3,\ya)(9,\yb)\rput(6,4.5){$\textstyle V_{2R}$}
\psline(2,9)(3,9)
\psline(2,6)(3,6)
%\psline(2,3)(3,3)
\psline(2,0)(3,0)
\psline(9,0)(10,0)
%\psline(9,3)(10,3)
\psline(9,6)(10,6)
\psline(9,9)(10,9)
\psline[linestyle=dotted](2,2)(2,4)
\psline[linestyle=dotted](10,2)(10,4)
%\psline[linestyle=dotted](11,2)(11,4)
\psset{origin={0,0}}
\rput[r](1.8,0){$\scriptstyle 1$}
\rput[r](1.8,6){$\scriptstyle R-1$}
\rput[r](1.8,9){$\scriptstyle R$}
\rput[l](10.2,9){$\scriptstyle R+1$}
\rput[l](10.2,6){$\scriptstyle R+2$}
\rput[l](10.2,0){$\scriptstyle 2R$}
\end{pspicture}}%
\end{center}
\caption{The composite operators $\O{}{R}$ \subref{corep} and
the Green function $V_{2R}$ \subref{Vrep} are the building blocks of the 
 $2$-point function $\big(\O{1}{R},V_{2R},\O{2}{R}\big)$}\label{bblockrep} 
\end{figure}
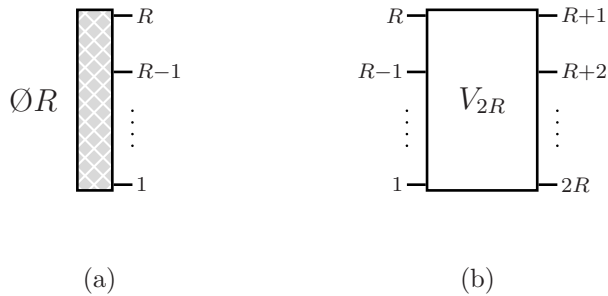

For the sake of simplicity, from now on we will use the abbreviation
\begin{equation}\label{traceabb}
\big(a_1\,a_2\,\dots\,a_n\big)=\tr(T^{a_1}T^{a_2}\dots T^{a_n})\col
\end{equation}
for the traces.
We represent $\O{}{R}$ by a rectangular box with $R$ elementary 
external legs attached to one of its sides. 
Counting of the lines starts with that line which with view into the outgoing
direction has no direct neighbour on its right hand side, and it ends with the
line that has no left hand neighbour, see Fig.\ \ref{corep}. 

By $V_{2R}$ we denote the Green functions of the theory with $2R$ 
elementary external legs. $R$ of them will be regarded as ingoing and 
outgoing, respectively. 
We represent $V_{2R}$ as a rectangular box where $R$ elementary 
field lines enter from the left hand side and the $R$ remaining lines from the
right hand side. Counting of the lines is clockwise and starts from the 
lower left corner, see Fig.\ \ref{Vrep}. 

The $2$-point function of two composite operators $\O{e}{R}$, is 
defined as the correlation function in which the $R$ legs of the operator 
$\O{1}{R}$ [$\O{2}{R}$] are 
entirely contracted with the $R$ ingoing [outgoing] fields of $V_{2R}$.
We denote such a $2$-point function by 
$\big(\O{1}{R},V_{2R},\O{2}{R}\big)$. 
When the two $\O{e}{R}$ and $V_{2R}$ are combined within a $2$-point function,
the elementary fields within the traces of $\O{e}{R}$ are counted with
increasing labels from left to right, starting with $\O{1}{R}$ from $1$ to
$R$, and continuing with $\O{2}{R}$ using the labels $R+1$ to $2R$. 
The $2$-point function is obtained by contracting the $2R$ fields of the
operators with the fields of $V_{2R}$ that carry the same label.
This procedure guarantees that there are no crossings between the connecting
lines if the lines leave the operator on the left [right] hand side from its  
right [left] hand side. 

\subsection{Coupling and genus expansion of $V_{2R}$}

\begin{figure}
\begin{center}
\begin{pspicture}(-18,-18)(18,7)%\showgrid
%\psframe(-18,-18)(18,7)
\setlength{\ya}{9\unit}
\addtolength{\ya}{0.5\doublesep}
\addtolength{\ya}{\linew}
\setlength{\yb}{0\unit}
\addtolength{\yb}{-0.5\doublesep}
\addtolength{\yb}{-\linew}
\sixdotsvertex{0}{0}{157.5}%{-22.5}
\psset{doubleline=true}
%e\psbezier(2.9;22.5)(5.226;22.5)(3,-1)(6,-1)
\psbezier(2.9;67.5)(12;67.5)(30,-15)(4,-15)
\psbezier(2.9;337.5)(9;337.5)(12,-9)(4,-9)
\psbezier(2.9;292.5)(4.6972;292.5)(7,-6)(4,-6)
%c\psbezier(2.9;112.5)(5.4119;112.5)(-3,5)(-6,5)
\psbezier(2.9;112.5)(12;112.5)(-30,-15)(-4,-15)
\psbezier(2.9;202.5)(9;202.5)(-12,-9)(-4,-9)
\psbezier(2.9;247.5)(4.6972;247.5)(-7,-6)(-4,-6)
\psset{doubleline=false}
\psline[linestyle=dotted](5,0)(5,2)
\psline[linestyle=dotted](-5,0)(-5,2)
\psline[linecolor=gray,linewidth=0.75\dlinewidth](0,-4)(0,4)
\psset{origin={6,15}}
\psframe(3,\ya)(9,\yb)\rput(0,-10.5){$\textstyle V_{2R}$}
\psset{doubleline=true}
\psline(2,9)(3,9)
\psline(2,6)(3,6)
%\psline(2,3)(3,3)
\psline(2,0)(3,0)
\psline(9,0)(10,0)
%\psline(9,3)(10,3)
\psline(9,6)(10,6)
\psline(9,9)(10,9)
\psset{doubleline=false}
%\psline[linestyle=dotted](2,2)(2,4)
%\psline[linestyle=dotted](10,2)(10,4)
\psline[linestyle=dotted](1,2)(1,4)
\psline[linestyle=dotted](11,2)(11,4)
\end{pspicture}
\end{center}
\caption{On a compact Riemann surface, the external lines of the Green
  function $V_{2R}$ are connected to the vertex at $\infty$. These connections
  separate the Riemann suface into two parts. Lines that cross these
  connections require adding of the wrapping handle, depicted in
  gray.}\label{GreenfuncplusVinfty}
\end{figure}
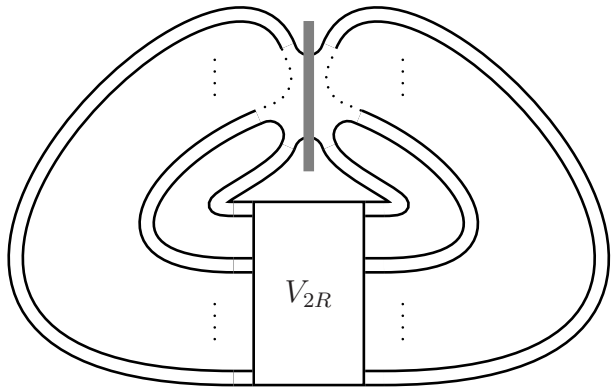

In perturbation theory, the Green function $V_{2R}$ can be regarded as 
a series in powers $K$ of the coupling constant $g$. Each fixed power $g^K$ 
contains a number of elementary Feynman diagrams. Denoting a particular
diagram by $D^K_{2R}$, one can write
\begin{equation}\label{couplingVexpand}
V_{2R}=\sum_K\sum_DD^K_{2R}\pnt
\end{equation}
In this expansion, any diagram $D^K_{2R}$ is of the order $g^K$.
Furthermore, one
can perform a genus expansion in powers of $N^{-2}$ \cite{'tHooft:1974jz}. 
Each Feynman diagram 
$D^K_{2R}$ is itself a sum of diagrams $\D{K}{h}{2R}$ which differ in the 
genus $h$. 
One defines the genus $h$ of a diagram as the minimal genus 
of all compact Riemannian surfaces on which the diagram can be drawn in
double-line notation without any crossings of lines. On the compactified 
Riemann surface, the $2R$ external lines
end at a common vertex (at the point representing $\infty$) with the reversed
ordering, i.e. they are attached counter-clockwise, starting with the lowest 
label. This configuration is shown in Fig.\ \ref{GreenfuncplusVinfty}.
Its genus can be obtained by Euler's relation for the Euler character $\chi$
\begin{equation}
\chi=2-2h=V-P+I\col
\end{equation}
where $V$ is the number of vertices, $P$ is the number of propagators, and
$I$ is the number of index loops.

The expansion of the Green function $V_{2R}$ then reads
\begin{equation}\label{genuscouplingVexpand}
V_{2R}=\sum_h\V{}{h}{2R}=\sum_h\sum_K\sum_D\D{K}{h}{2R}\col
\end{equation}
where each contribution $\V{}{h}{2R}$ with fixed genus $h$ is of order
$N^{2-2h}$, and each diagram $\D{K}{h}{2R}$ is of the order
$\sqrt{\lambda}^KN^{2-2h}$. 

The vertex at $\infty$ carries a normalization factor $N^{1-R}$, which can
be understood as follows.
The vertex at $\infty$ can be regarded as an effective vertex 
for a planar connected ($C=1$) tree-level ($L=0$) diagram with 
$E=2R$ external legs. According to \eqref{E} it can be built out of $2R-2$ 
$3$-vertices. Hence, it should depend on $\sqrt{\lambda}^{2R-2}$. The
normalization factor is obtained by providing the required $N$ dependence 
to transform the YM coupling $g$ into $\lambda$, and then setting $\lambda=1$
for the effective vertex.

\subsection{Coupling and genus expansion of the $2$-point function}
\label{2ptexpansion}
Similarly to $V_{2R}$, one can expand the $2$-point function
$\big(\O{1}{R},V_{2R},\O{2}{R}\big)$ in powers of $g$ and $N^{-2}$.
A single diagram $\D{K}{h}{2R}$ generates a single diagram of the $2$-point
function if all $(R!)^2$ permutations within the two bundles of $R$ elementary
legs are contained within $V_{2R}$. This means each permutation of the 
external legs is regarded as a different diagram $\D{K}{h}{2R}$. Then, one only
has to consider 
a unique way to contract the $2R$ legs of the two operators $\O{e}{R}$ with
the external legs of $\D{K}{h}{2R}$.

Consider the genus expansion of the $2$-point function
$\big(\O{1}{R},V_{2R},\O{2}{R}\big)$. 
By $\big(\O{1}{R},\D{K}{h}{2R},\O{2}{R}\big)^{(H)}$ we denote the genus $H$
contribution of a particular diagram $\D{K}{h}{2R}$ to the $2$-point function.
The genus $h$ of a particular diagram $\D{K}{h}{2R}$ does \emph{not} 
uniquely determine the genus $H$. This effect has two reasons.

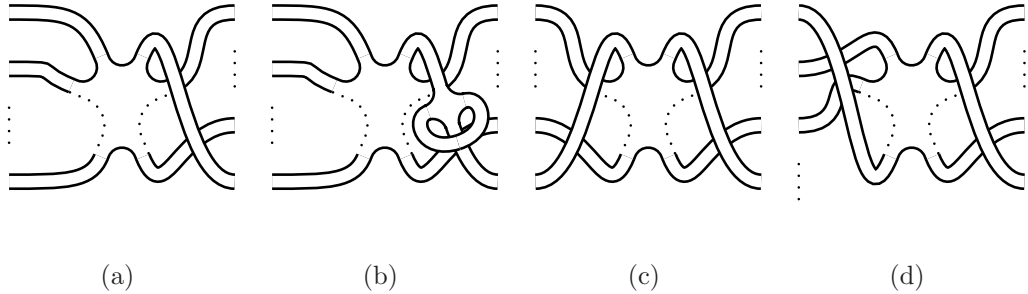
\begin{figure}
\begin{center}%
%\subfigure[]{\label{planarVinfty}%
%\begin{pspicture}(-7,-6)(7,7)%\showgrid
%\psframe(-7,-6)(7,7)
%\sixdotsvertex{0}{0}{-22.5}
%\psset{doubleline=true}
%\psbezier(2.9;292.5)(4.3295;292.5)(3,-4)(6,-4)
%\psbezier(2.9;22.5)(5.226;22.5)(3,2)(6,2)
%\psbezier(2.9;67.5)(5.4119;67.5)(3,5)(6,5)
%\psbezier(2.9;112.5)(5.4119;112.5)(-3,5)(-6,5)
%\psbezier(2.9;157.5)(5.226;157.5)(-3,2)(-6,2)
%\psbezier(2.9;247.5)(4.3295;247.5)(-3,-4)(-6,-4)
%\psset{doubleline=false}
%\psline[linestyle=dotted](6,-2)(6,0)
%\psline[linestyle=dotted](-6,-2)(-6,0)
%\end{pspicture}}%
\subfigure[]{\label{rcyclVinfty}%
\begin{pspicture}(-7,-6)(7,7)%\showgrid
%\psframe(-7,-6)(7,7)
\sixdotsvertex{0}{0}{-22.5}
\psset{doubleline=true}
\psbezier(2.9;292.5)(6;292.5)(3,-1)(6,-1)
\psbezier(2.9;22.5)(5.226;22.5)(3,5)(6,5)
\psbezier(2.9;67.5)(7;67.5)(3,-4)(6,-4)
\psbezier(2.9;112.5)(5.4119;112.5)(-3,5)(-6,5)
\psbezier(2.9;157.5)(5.226;157.5)(-3,2)(-6,2)
\psbezier(2.9;247.5)(4.3295;247.5)(-3,-4)(-6,-4)
\psset{doubleline=false}
\psline[linestyle=dotted](6,1)(6,3)
\psline[linestyle=dotted](-6,-2)(-6,0)
\end{pspicture}}%
\subfigure[]{\label{rhandelcyclVinfty}%
\begin{pspicture}(-7,-6)(7,7)%\showgrid
%\psframe(-7,-6)(7,7)
\sixdotsvertex{0}{0}{-22.5}
\psset{doubleline=true}
\psbezier(2.9;292.5)(6;292.5)(3,-1)(6,-1)
\psbezier(2.9;22.5)(5.226;22.5)(3,5)(6,5)
\psbezier(2.9;67.5)(7;67.5)(3,-4)(6,-4)
\psbezier(2.9;112.5)(5.4119;112.5)(-3,5)(-6,5)
\psbezier(2.9;157.5)(5.226;157.5)(-3,2)(-6,2)
\psbezier(2.9;247.5)(4.3295;247.5)(-3,-4)(-6,-4)
\psset{doubleline=false}
\psline[linestyle=dotted](6,1)(6,3)
\psline[linestyle=dotted](-6,-2)(-6,0)
\fourvertex{3.27}{0}{17.75}
\psset{doubleline=true}
% rotate by -72.25 with origin (3.27,0)
%\psbezier[doubleline=true](0,1)(0,2)(2,2)(2,0)
%\psbezier[liftpen=2](0,-1)(0,-2)(2,-2)(2,0)
\psset{origin={-3.27,0}}
\psbezier[doubleline=true](1;17.75)(2;17.75)(2.8284;-27.25)(2;-72.25)
\psbezier[doubleline=true](1;197.75)(2;197.75)(2.8284;-117.25)(2;-72.25)
\end{pspicture}}%
\subfigure[]{\label{rlcyclVinfty1}%
\begin{pspicture}(-7,-6)(7,7)%\showgrid
%\psframe(-7,-6)(7,7)
\sixdotsvertex{0}{0}{-22.5}
\psset{doubleline=true}
\psbezier(2.9;292.5)(6;292.5)(3,-1)(6,-1)
\psbezier(2.9;22.5)(5.226;22.5)(3,5)(6,5)
\psbezier(2.9;67.5)(7;67.5)(3,-4)(6,-4)
\psbezier(2.9;157.5)(5.226;157.5)(-3,5)(-6,5)
\psbezier(2.9;247.5)(6;247.5)(-3,-1)(-6,-1)
\psbezier(2.9;112.5)(7;112.5)(-3,-4)(-6,-4)
\psset{doubleline=false}
\psline[linestyle=dotted](6,1)(6,3)
\psline[linestyle=dotted](-6,1)(-6,3)
\end{pspicture}}%
\subfigure[]{\label{rlcyclVinfty2}%
\begin{pspicture}(-7,-6)(7,7)%\showgrid
%\psframe(-7,-6)(7,7)
\sixdotsvertex{0}{0}{-22.5}
\psset{doubleline=true}
\psbezier(2.9;292.5)(6;292.5)(3,-1)(6,-1)
\psbezier(2.9;22.5)(5.226;22.5)(3,5)(6,5)
\psbezier(2.9;67.5)(7;67.5)(3,-4)(6,-4)
\psbezier(2.9;112.5)(5.4119;112.5)(-3,2)(-6,2)
\psbezier(2.9;157.5)(5.226;157.5)(-3,-1)(-6,-1)
\psbezier(2.9;247.5)(8;247.5)(-3,5)(-6,5)
\psset{doubleline=false}
\psline[linestyle=dotted](6,1)(6,3)
\psline[linestyle=dotted](-6,-5)(-6,-3)
\end{pspicture}}%
\end{center}
\caption{The cyclic permutations of the
    external legs have been moved to the vertex at $\infty$.
The resolution of the cyclic permutation 
    requires adding one handle \subref{rcyclVinfty}.
An exception is the case \subref{rhandelcyclVinfty} in which 
the crossing leg contains a non-planar self energy correction. In this case
the genus does not change.
Two cyclic permutations in the same 
\subref{rlcyclVinfty1} or in opposite directions \subref{rlcyclVinfty2} can be
    resolved with a single handle.}
\label{Vinfty}
\end{figure}

\begin{figure}
\begin{center}
\begin{pspicture}(-18,-18)(18,7)%\showgrid
%\psframe(-18,-18)(18,7)
\psset{origin={0,3}}
\drinsert{-1}{0}
\orvertex{-1}{3}
\ordots{-1}{6}
\urinsert{-1}{9}
\ulinsert{1}{9}
\oldots{1}{6}
\olvertex{1}{3}
\dlinsert{1}{0}
%\psset{origin={0,0}}
%\sixdotsvertex{0}{0}{157.5}%{-22.5}
\psset{doubleline=true}
%e\psbezier(2.9;22.5)(5.226;22.5)(3,-1)(6,-1)
\psbezier(-4,0)(-6,0)(-6,-3)(-4,-3)
\psbezier(-4,3)(-9,3)(-9,-6)(-4,-6)
\psbezier(-4,9)(-15,9)(-15,-12)(-4,-12)
\psbezier(4,0)(6,0)(6,-3)(4,-3)
\psbezier(4,3)(9,3)(9,-6)(4,-6)
\psbezier(4,9)(15,9)(15,-12)(4,-12)
%\psbezier(2.9;337.5)(9;337.5)(12,-9)(4,-9)
%\psbezier(2.9;292.5)(4.6972;292.5)(7,-6)(4,-6)
%c\psbezier(2.9;112.5)(5.4119;112.5)(-3,5)(-6,5)
%\psbezier(2.9;112.5)(12;112.5)(-30,-15)(-4,-15)
%\psbezier(2.9;202.5)(9;202.5)(-12,-9)(-4,-9)
%\psbezier(2.9;247.5)(4.6972;247.5)(-7,-6)(-4,-6)
\psset{doubleline=false}
\psline[linestyle=dotted](5,5)(5,7)
\psline[linestyle=dotted](-5,5)(-5,7)
\psline[linecolor=gray,linewidth=0.75\dlinewidth](0,-1)(0,10)
\setlength{\ya}{9\unit}
\addtolength{\ya}{0.5\doublesep}
\addtolength{\ya}{\linew}
\setlength{\yb}{0\unit}
\addtolength{\yb}{-0.5\doublesep}
\addtolength{\yb}{-\linew}
\psset{origin={6,15}}
\psframe(3,\ya)(9,\yb)\rput(0,-10.5){$\textstyle V_{2R}$}
\psset{doubleline=true}
\psline(2,9)(3,9)
\psline(2,6)(3,6)
%\psline(2,3)(3,3)
\psline(2,0)(3,0)
\psline(9,0)(10,0)
%\psline(9,3)(10,3)
\psline(9,6)(10,6)
\psline(9,9)(10,9)
\psset{doubleline=false}
%\psline[linestyle=dotted](2,2)(2,4)
%\psline[linestyle=dotted](10,2)(10,4)
\psline[linestyle=dotted](1,2)(1,4)
\psline[linestyle=dotted](11,2)(11,4)
\end{pspicture}
\end{center}
\caption{On a compact Riemann surface, the two operators $\O{e}{R}$ in the
  $2$-point function have a finite separation. This allows one to occupy the 
  wrapping path, depicted in   
  gray.}\label{2ptfunc}
\end{figure}
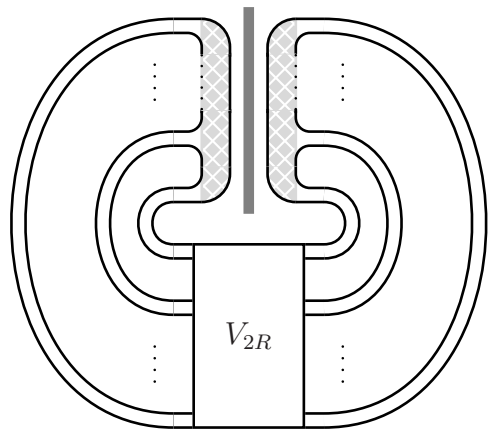

First of all, drawing a particular diagram of the $2$-point function, one 
has drawn only one representative of an $R^2$-dimensional equivalence class of 
diagrams. It contains all diagrams that differ only by cyclic permutations 
within each of the two pairs of legs that connect the diagram $\D{K}{h}{2R}$
to the two operators $\O{e}{2R}$. 
The equivalence relation is based on the fact that the
operators \eqref{codef} are invariant under cyclic permutations of their 
elementary fields.
Two different representations contain two different diagrams $\D{K}{h}{2R}$
and $\D{K}{h'}{2R}$ that can be mapped to each other 
by acting with cyclic permutations on the ingoing and/or outgoing legs. 
All these diagrams are again members of an equivalence class, for which
elements one has $h'\neq h$ in general.
Let us denote by $h$ the minimal genus of all the diagrams within one
equivalence class.
A cyclic permutation applied to either the incoming or outgoing legs increases
the genus $h$ by at most one. This can be seen by resolving the 
crossing at the vertex at $\infty$, see Fig.\ \ref{rcyclVinfty}. 
It requires adding at most one handle. In case that the crossing line contains
a non-planar self energy contribution, no futher handle has to be
added. A further cyclic permutation acting
on the remaining $R$ legs does \emph{not} change the genus further if 
the cyclic permutation applied to the other $R$ legs has already
increased the genus. This is because 
the same handle can be used to resolve the crossing, see Fig.\
\ref{rlcyclVinfty1} and \ref{rlcyclVinfty2}.
However, in case that no handle had to be added for the first set of 
$R$ legs (like in case of a crossing line with a non-planar self
energy contribution), at most one handle has to be added for the
cyclic permutation applied to the second set of $R$ fields.

%\begin{figure}
%\begin{center}
%\rule[-4cm]{0pt}{4cm}
%\end{center}
%\caption{A handle can resolve a crossing between two lines that run along it.}
%\label{crossingonhandle}
%\end{figure}

Under the $R^2$
diagrams $\D{K}{h'}{2R}$ there is at least one with minimal genus $h'=h$. The
remaining ones have genus $h\le h'\le
  h+1$ because the same handle can be used to resolve the two
  crossings caused by both cyclic permutations.
%\footnote{\color{red} That the difference between $h'$ and $h$
%  is at 
%  most one is due to the fact that each handle allows one to resolve
%  one crossing between lines that run along this handle, see Fig.\
%  \ref{crossingonhandle}.}
This effect means that in 
$\big(\O{1}{R},\D{K}{h}{2R},\O{2}{R}\big)^{(H)}$ the genus $H$ can have values 
$H=h-1$ or $H=h$.

The second reason for the genus expansion of $V_{2R}$ and of
$\big(\O{1}{R},V_{2R},\O{2}{R}\big)$ to differ from each other is
caused by the fact that the $2$-point function  
does not have external lines which, when it is drawn on a compact
Riemann surface, have to be connected to an additional 
vertex at $\infty$. 
The bundles of ingoing and outgoing lines, interacting with each other 
in $V_{2R}$ and in the vertex at $\infty$, form a closed contour 
which divides the Riemann surface into two parts, see Fig.\
\ref{GreenfuncplusVinfty}. 
That means, there has to be one handle for
all the field lines crossing this contour. We will call this handle the 
wrapping handle. While being needed in certain diagrams $\D{K}{h}{2R}$, 
it can be removed in 
$\big(\O{1}{R},\D{K}{h}{2R},\O{2}{R}\big)$. Due to the separation of the two
operators $\O{e}{R}$ their contraction with the external lines of $V_{2R}$ no
longer forms a closed contour, that divides the Riemann surface into two
parts. There is a direct connection, from now 
on called the wrapping path, that replaces the wrapping handle
(compare Fig.\ \ref{GreenfuncplusVinfty} and Fig.\ \ref{2ptfunc}).  
The genus $H$ of $\big(\O{1}{R},\D{K}{h}{2R},\O{2}{R}\big)$ hence obeys 
$h-1\le H\le h$.
Combining the above described effects, one expects that the diagrams 
$\big(\O{1}{R},\D{K}{h}{2R},\O{2}{R}\big)^{(H)}$ obtained from a genus $h$
diagram $\D{K}{h}{2R}$ in general have $h-2\le H\le h$.

\begin{figure}
\begin{center}
\begin{pspicture}(-4,-2)(38,11)%\showgrid
%\psframe(-4,-2)(38,11)
\setlength{\ya}{9\unit}
\addtolength{\ya}{0.5\doublesep}
\addtolength{\ya}{\linew}
\setlength{\yb}{0\unit}
\addtolength{\yb}{-0.5\doublesep}
\addtolength{\yb}{-\linew}
\psframe[fillstyle=\olfillstyle,fillcolor=\olcolor,hatchcolor=\olhatchcolor,
%linecolor=\olcolor,
linewidth=\linew](0,\ya)(2,\yb)\rput[r](-1,4.5){$\textstyle C_R\big(\O{1}{R}\big)$}
\psline(2,9)(3,9)
\psline(2,6)(3,6)
%\psline(2,3)(3,3)
\psline(2,0)(3,0)
%\psline[linestyle=dotted](3,2)(3,4)
\psline[linestyle=dotted](4,2)(4,4)
\psset{origin={-2,0}}
\psframe(4,\ya)(8,\yb)\rput(8,4.5){$\textstyle\frac{S_R}{C_R}$}
\psline(3,9)(4,9)
\psline(3,6)(4,6)
%\psline(3,3)(4,3)
\psline(3,0)(4,0)
\psline(8,0)(9,0)
%\psline(8,3)(9,3)
\psline(8,6)(9,6)
\psline(8,9)(9,9)
%\psline[linestyle=dotted](3,2)(3,4)
%\psline[linestyle=dotted](9,2)(9,4)
\psline[linestyle=dotted](10,2)(10,4)
\psset{origin={-11,0}}
\psframe(3,\ya)(9,\yb)\rput(17,4.5){$\textstyle\frac{V_{2R}}{S_R}$}
\psline(2,9)(3,9)
\psline(2,6)(3,6)
%\psline(2,3)(3,3)
\psline(2,0)(3,0)
\psline(9,0)(10,0)
%\psline(9,3)(10,3)
\psline(9,6)(10,6)
\psline(9,9)(10,9)
%\psline[linestyle=dotted](2,2)(2,4)
%\psline[linestyle=dotted](10,2)(10,4)
\psline[linestyle=dotted](11,2)(11,4)
\psset{origin={-20,0}}
\psframe(4,\ya)(8,\yb)\rput(26,4.5){$\textstyle\frac{S_R}{C_R}$}
\psline(3,9)(4,9)
\psline(3,6)(4,6)
%\psline(3,3)(4,3)
\psline(3,0)(4,0)
\psline(8,0)(9,0)
%\psline(8,3)(9,3)
\psline(8,6)(9,6)
\psline(8,9)(9,9)
%\psline[linestyle=dotted](3,2)(3,4)
%\psline[linestyle=dotted](9,2)(9,4)
\psline[linestyle=dotted](10,2)(10,4)
\psset{origin={-32,0}}
\psframe[fillstyle=\orfillstyle,fillcolor=\orcolor,hatchcolor=\orhatchcolor,
%linecolor=\orcolor,
linewidth=\linew](0,\ya)(2,\yb)\rput[l](35,4.5){$\textstyle C_R\big(\O{2}{R}\big)$}
\psline(-1,9)(0,9)
\psline(-1,6)(0,6)
%\psline(-1,3)(0,3)
\psline(-1,0)(0,0)
%\psline[linestyle=dotted](-1,2)(-1,4)
%\psline[linestyle=dotted](-2,2)(-2,4)
%
\psset{origin={0,0}}
\psline(3,0)(5,0)\rput[b](4,0.2){$\scriptstyle 1$}
\psline(3,6)(5,6)\rput[b](4,6.2){$\scriptstyle R-1$}
\psline(3,9)(5,9)\rput[b](4,9.2){$\scriptstyle R$}
\psline(11,0)(13,0)
\psline(11,6)(13,6)
\psline(11,9)(13,9)
\psline(21,9)(23,9)
\psline(21,6)(23,6)
\psline(21,0)(23,0)
\psline(29,9)(31,9)\rput[b](30,9.2){$\scriptstyle R+1$}
\psline(29,6)(31,6)\rput[b](30,6.2){$\scriptstyle R+2$}
\psline(29,0)(31,0)\rput[b](30,0.2){$\scriptstyle 2R$}
\rput[t](17,-1){$\underbrace{\rule[0pt]{24\unit}{0pt}}_{\displaystyle V_{2R}}$}
\end{pspicture}
\end{center}
\caption{
Decomposition of the $2$-point function
 $\big(\O{1}{R},V_{2R},\O{2}{R}\big)$ into interaction part
 $\frac{V_{2R}}{S_R}$ without any permutations of the external legs and the 
permutations $\frac{S_R}{C_R}$ without cyclic permutations $C_R$. 
$V_{2R}$ contains only one diagram $\D{K}{h}{2R}$ 
of each equivalence class with minimal genus $h$.
The cyclic permutations are taken into account by using all cyclic
permutations of the operators $\O{e}{R}$, denoted by $C_R\big(\O{e}{R}\big)$.}
\label{2ptfunctiondecomp}
\end{figure}
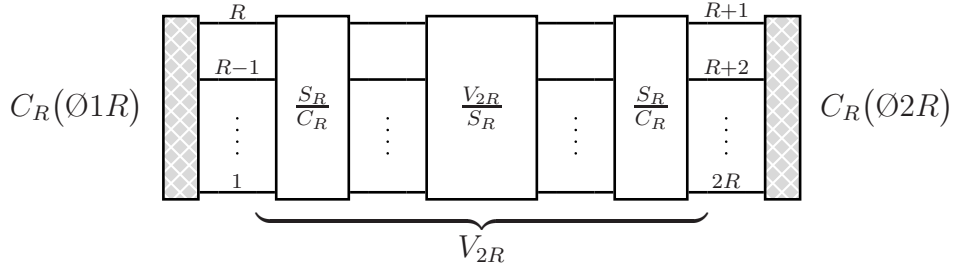

Since here we are interested only in the change of the genus caused by 
removing the wrapping handle, we will disentangle both effect.
From each equivalence class of diagrams we will only use one representative
$\D{K}{h}{2R}$ with the minimal genus $h$ to build the corresponding
diagram of the $2$-point function. 
The additonal $R^2-1$ diagrams generated by cyclic permutations 
are then taken care of by the two operators $\O{e}{R}$. 
That means, to each of the operators we associate $R-1$ 
copies that differ by cyclic permutations. They are combined in $R^2$ 
possible ways to obtain all contributions to the $2$-point function that
contain the diagrams $\D{K}{h'}{2R}$ with $h\le h'\le h+1$ in the equivalence
class of $\D{K}{h}{2R}$. The distribution of the 
permutations of the external legs of $V_{2R}$ is depicted in Fig.\
\ref{2ptfunctiondecomp}. 
    
After that we are left with two types of genus $h$ diagrams 
$\D{K}{h}{2R}$. One type of $\D{K}{h}{2R}$ contributes to genus $h$ diagrams 
of the $2$-point function $\big(\O{1}{R},\D{K}{h}{2R},\O{2}{R}\big)$, the
other type contributes to genus $h-1$ diagrams, and its elements are called 
wrapping diagrams. In the next Section we will analyze these 
contributions in detail. 

\section{Wrapping diagrams}
\label{wrapdiag}
\subsection{Definition of wrapping and non-wrapping diagrams}
\begin{figure}
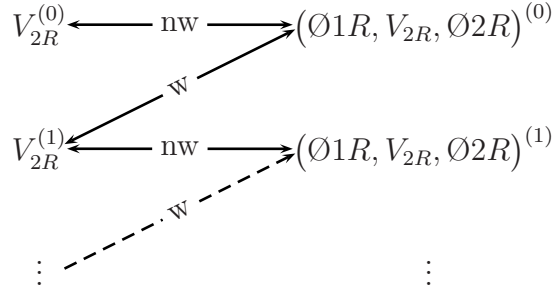

\begin{center}
\begin{tabular}{@{}c@{}@{}m{3cm}@{}@{}c@{}}
\rule{0pt}{1.5cm}\hfill\rnode[r]{n11}{$\V{}{0}{2R}$} & & \rnode[l]{n12}{$\big(\O{1}{R},V_{2R},\O{2}{R}\big)^{(0)}$}
\hspace*{\fill} \\
\rule{0pt}{1.5cm}\hfill\rnode[r]{n21}{$\V{}{1}{2R}$} & & \rnode[l]{n22}{$\big(\O{1}{R},V_{2R},\O{2}{R}\big)^{(1)}$}
\hspace*{\fill}  \\
\rule{0pt}{1.5cm}\hspace*{\fill}$\vdots$\hfill\rnode[r]{n}{$\vphantom{\big(\O{1}{R},V_{2R},\O{2}{R}\big)^{(h)}}$} & &\rnode[l]{m}{$\vphantom{\big(\O{1}{R},V_{2R},\O{2}{R}\big)^{(h+1)}}$}\hfill$\vdots$\hspace*{\fill} \\
%\rule{0pt}{1.5cm}\hfill\rnode[r]{n31}{$\V{}{h}{2R}$} & & \rnode[l]{n32}{$\big(\O{1}{R},V_{2R},\O{2}{R}\big)^{(h)}$}
%\hspace*{\fill}  \\
%\rule{0pt}{1.5cm}\hfill\rnode[r]{n41}{$\V{}{h+1}{2R}$} & & \rnode[l]{n42}{$\big(\O{1}{R},V_{2R},\O{2}{R}\big)^{(h+1)}$}
%\hspace*{\fill} \\
%\rule{0pt}{1.5cm}\hspace*{\fill}$\vdots$\hfill\rnode[r]{r}{$\vphantom{\big(\O{1}{R},V_{2R},\O{2}{R}\big)^{(h)}}$} & &\rnode[l]{s}{$\vphantom{\big(\O{1}{R},V_{2R},\O{2}{R}\big)^{(h+1)}}$}\hfill$\vdots$\hspace*{\fill} \\
\end{tabular}
\psset{arrows=<->}
\ncline{n11}{n12}\lput*{:U}{nw}
\ncline[offsetA=2pt,offsetB=-2pt]{n21}{n12}\lput*{:U}{w}
\ncline{n21}{n22}\lput*{:U}{nw}
\ncline{n31}{n32}\lput*{:U}{nw}
\ncline[offsetA=2pt,offsetB=-2pt]{n41}{n32}\lput*{:U}{w}
\ncline{n41}{n42}\lput*{:U}{nw}
\ncline[offsetA=2pt,offsetB=-2pt,linestyle=dashed,arrows=->]{n}{n22}\lput*{:U}{w}
\ncline[offsetA=2pt,offsetB=-2pt,linestyle=dashed,arrows=<-]{n31}{m}\lput*{:U}{w}
\psset{arrows=-}
\ncline[offsetA=2pt,offsetB=-2pt,linestyle=dashed,arrows=->]{r}{n42}\lput*{:U}{w}
\caption{The translation of the genus expansion between $\V{}{h}{2R}$ and
$\big(\O{1}{R},V_{2R},\O{2}{R}\big)^{(H)}$. 
The non wrapping ($\text{nw}$) genus $h$ contributions $\Vnw{}{h}{2R}$ 
to the Green function become genus $H=h$ contributions
$\big(\O{1}{R},V_{2R},\O{2}{R}\big)^{(H)}$ to the $2$-point function. 
The wrapping ($\text{w}$) contributions 
$\Vw{}{h}{2R}$ become genus $H=h-1$ contributions to the $2$-point function.}
\label{Vversus2ptgenusexpand}
\end{center}
\end{figure}

We define the \emph{genus $h-1$ wrapping diagrams} $\Dw{K}{h}{2R}$ as the genus
$h$ contributions to $V_{2R}$ that lead to a genus $H=h-1$ contribution 
to the $2$-point function $\big(\O{1}{R},V_{2R},\O{2}{R}\big)$.
Correspondingly, the non-wrapping diagrams $\Dnw{K}{h}{2R}$ are the remaining
genus $h$ diagrams that generate genus $H=h$ contributions 
to $\big(\O{1}{R},V_{2R},\O{2}{R}\big)$.
One has the relations
\begin{equation}\label{wrapnonwrapVdecomp}
\V{}{h}{2R}=\Vnw{}{h}{2R}+\Vw{}{h}{2R}
\end{equation}
\begin{equation}\label{wrapnonwrap2ptdecomp}
\begin{aligned}
\big(\O{1}{R},\V{}{h}{2R},\O{2}{R}\big)&=\big(\O{1}{R},\Vnw{}{h}{2R},\O{2}{R}\big)^{(h)}+\big(\O{1}{R},\Vw{}{h}{2R},\O{2}{R}\big)^{(h-1)}\col\\
 \big(\O{1}{R},V_{2R},\O{2}{R}\big)^{(H)}&=\big(\O{1}{R},\Vnw{}{H}{2R},\O{2}{R}\big)+\big(\O{1}{R},\Vw{}{H+1}{2R},\O{2}{R}\big)\pnt
\end{aligned}
\end{equation}
The translation of the genus expansion between $\V{}{h}{2R}$ and
$\big(\O{1}{R},V_{2R},\O{2}{R}\big)^{(H)}$ is visualized in Fig.\ \ref{Vversus2ptgenusexpand}.

\subsection{Spectator fields}
\label{spectatorfields}
Here we develop a useful tool 
to distinguish between planar wrapping and non wrapping diagrams 
based on adding spectator fields. 
They come in pairs of two fields $\psi$, $\hat \psi$ that are inserted into
the trace of respectively $\O{1}{2R}$ and $\O{2}{2R}$, defined in
\eqref{codef}, at any position between two $\phi_{n_r}$.
The contraction of a pair of spectator fields, denoted by 
$\cpair[0.1cm]{\psi}{\hat\psi}$,  generates a further connection
line within the diagram, not interacting with any other lines.
Pairs of spectator fields must be added such that their connection lines 
do not cross each other, and such that not two pairs are
equivalent. Two pairs are equivalent if both, 
the two $\psi$ and the $\hat\psi$ are direct neighbours within the two
traces, not being separated by at least one $\phi_{n_r}$.    
The spectator pairs can be used to test 
some aspects of the topology of a diagram by observing how the  
genus $H$ of the $2$-point function behaves when their connection lines are 
added. 

In contrast to the wrapping diagrams, for non-wrapping ones there exists at
least one representation in which 
the wrapping path introduced in Subsection \ref{2ptexpansion} is unoccupied. 
Hence, following this empty path,
it is possible to add a pair of spectator fields 
such that their connection does not cross any of the other lines.

Due to the fact that the lines along the wrapping path differ between two
representations of the same diagram, it is 
\emph{not sufficient} 
for a diagram to be wrapping that its wrapping path is 
occupied by at least one field line. Likewise, it is 
\emph{not necessary} 
for a diagram to be non-wrapping that its wrapping path is unoccupied.
%The symmetry of the operators \eqref{codef} under cyclic
%permutations allows one to change the representation of a given diagram.
%The wrapping path changes in each of these equivalent representations.
Instead,
to distinguish between a wrapping and a non-wrapping diagram, one has to
ensure that none or respectively at least one equivalent diagram with an
unoccupied wrapping path exists.

Consider a candidate for a genus $h-1$ wrapping diagram.
In a given representation of such a diagram one can always free the 
wrapping path by removing some lines. No further lines that have no 
influence on the occupation of the wrapping path are removed.
In a second step one removes the 
two operators $\O{e}{2R}$. In this way one arrives at a non-wrapping
diagram $\Dnw{K'}{h'}{2R}$ that is a truncated version of the original diagram 
$\D{K}{h}{2R}$, and that obeys $K'\le K$, $h'\le h-1$.
If the original diagram $\D{K}{h}{2R}$ was wrapping, then the
truncated diagram $\Dnw{K'}{h'}{2R}$ is necessarily connected ($C=1$
in the notation introduced in Appendix \ref{countingrules}, 
see \eqref{diagquantities}). This statement is only true  
if one forbids that $\D{K}{h}{2R}$ contains connected pieces
that are entirely attached to only one of the two sets of $R$ external legs.
We show that if $C>1$ under the above assumptions, the original diagram was
non-wrapping.  
If $C>1$, one has two possibilities to add two distinct spectator lines
without introducing new handles or crossings, 
that separate the two connected pieces. One of them obstructs the wrapping
path. It is possible to add these spectator lines such that they connect the
operators with each other. 
This is possible
since we demand that each connected piece was connected to both
operators.
Restoring the wrapping lines would force one to 
remove one of the spectator lines, since otherwise it had to be crossed, 
but the other spectator line can be kept, and hence the diagram would be
non-wrapping for $C>1$. 

In Appendix \ref{countingrules} we show that for a theory with
vertices with $k+2$ legs that 
are of order $g^k$ in the YM coupling constant, planar wrapping
interactions of order $K$ in the YM coupling constant exist for
$K\geq2R$. As long as  
$K\leq2(2R-1)$ they can be constructed from non-wrapping diagrams by
adding a single line  that runs along the wrapping path. 
This observation might be useful in explicit computer based
calculations like \cite{Fischbacher:2004iu}, putting the analysis at
higher order ($K\leq2(2R-1)$) on an equal footing with the critical
case ($K=2R$).

\subsection{The encoding of the topology of planar wrapping diagrams by
  the spectator fields}
\label{topencoding} 

\begin{figure}
\begin{center}
\subfigure[]{\label{genus2nonwrap}%
\begin{pspicture}(-1,-4)(13,13)%\showgrid
%\psframe(-1,-4)(13,13)
\ulinsert[sgray]{0}{9}
\olvertex{0}{6}
\olvertex[sgray]{0}{3}
\dlinsert{0}{0}
\drinsert{12}{0}
\orvertex[sgray]{12}{3}
\orvertex{12}{6}
\urinsert[sgray]{12}{9}
\psset{doubleline=true}
\psline[linearc=2\linearc,doublecolor=sgray](3,9)(9,9)
\psline[linearc=2\linearc,doublecolor=sgray](3,3)(9,3)
%\psline[linearc=2\linearc](8,7)(8,12)(15,12)(15,-3)(6,-3)(6,-1)
%\psline(6,5)(6,1)
\psline(8,5)(8,1)
\threevertex{4}{6}{90}
\threevertex{6}{6}{90}
\threevertex{8}{6}{-90}
\threevertex{4}{0}{90}
\threevertex{6}{0}{90}
\threevertex{8}{0}{90}
\psset{doubleline=true}
\psbezier(4,7)(4,8)(3,8)(3,7)
\psbezier(3,7)(3,4)(6,4)(6,1)
\psbezier(6,7)(6,8)(7,8)(7,7)
\psbezier(7,7)(7,4)(4,4)(4,1)
%\psbezier(4,5)(4,3)(8,3)(8,1)
%\psbezier(4,1)(4,3)(8,3)(8,5)\psline(6,5)(6,1)
\end{pspicture}}%
\begin{pspicture}(0,0)(3.5,0)
\end{pspicture}
\subfigure[]{\label{genus1wrap1}%
\begin{pspicture}(-1,-4)(16.5,13)%\showgrid
%\psframe(-1,-4)(16.5,13)
\ulinsert[sgray]{0}{9}
\olvertex{0}{6}
\olvertex[sgray]{0}{3}
\dlinsert{0}{0}
\drinsert{12}{0}
\orvertex[sgray]{12}{3}
\orvertex{12}{6}
\urinsert[sgray]{12}{9}
\psset{doubleline=true}
\psline[linearc=2\linearc,doublecolor=sgray](3,9)(9,9)
\psline[linearc=2\linearc,doublecolor=sgray](3,3)(9,3)
%\psline[linearc=2\linearc](8,7)(8,12)(15,12)(15,-3)(6,-3)(6,-1)
%\psline(6,5)(6,1)
\psline(8,5)(8,1)
\threevertex{4}{6}{90}
\threevertex{6}{6}{-90}
\threevertex{8}{6}{-90}
\threevertex{4}{0}{-90}
\threevertex{6}{0}{90}
\threevertex{8}{0}{90}
\psset{doubleline=true}
\psbezier(4,7)(4,8)(3,8)(3,7)
\psbezier(3,7)(3,4)(6,4)(6,1)
\psline[linearc=2\linearc](7,5)(7,12)(15,12)(15,-3)(4,-3)(4,-1)
\psbezier(7,5)(7,4)(6,4)(6,5)
%\psbezier(6,5)(6,3)(3,3)(3,-1)
%\psbezier(3,-1)(3,-2)(4,-2)(4,-1)
%\psbezier(4,5)(4,3)(8,3)(8,1)
%\psbezier(4,1)(4,3)(8,3)(8,5)\psline(6,5)(6,1)
\end{pspicture}}%
\subfigure[]{\label{genus1wrap2}%
\begin{pspicture}(-1,-4)(16.5,13)%\showgrid
%\psframe(-1,-4)(16.5,13)
\ulinsert[sgray]{0}{9}
\olvertex{0}{6}
\olvertex[sgray]{0}{3}
\dlinsert{0}{0}
\drinsert{12}{0}
\orvertex[sgray]{12}{3}
\orvertex{12}{6}
\urinsert[sgray]{12}{9}
\psset{doubleline=true}
\psline[linearc=2\linearc,doublecolor=sgray](3,9)(9,9)
\psline[linearc=2\linearc,doublecolor=sgray](3,3)(9,3)
\psline[linearc=2\linearc](8,7)(8,12)(15,12)(15,-3)(8,-3)(8,-1)
\threevertex{4}{6}{-90}
\threevertex{6}{6}{-90}
\threevertex{8}{6}{90}
\threevertex{4}{0}{90}
\threevertex{6}{0}{90}
\threevertex{8}{0}{-90}
\psset{doubleline=true}
\psbezier(4,5)(4,3)(6,3)(6,1)
\psbezier(4,1)(4,3)(6,3)(6,5)
\end{pspicture}}\\%
\subfigure[]{\label{genus2wrap}
\begin{pspicture}(-1,-4)(16,19)%\showgrid
%\psframe(-1,-4)(16,19)
\ulinsert[sgray]{0}{15}
\olvertex{0}{12}
\olvertex[sgray]{0}{9}
\olvertex{0}{6}
\olvertex[sgray]{0}{3}
\dlinsert{0}{0}
\drinsert{12}{0}
\orvertex[sgray]{12}{3}
\orvertex{12}{6}
\orvertex[sgray]{12}{9}
\orvertex{12}{12}
\urinsert[sgray]{12}{15}
\psset{doubleline=true}
\psline[linearc=2\linearc,doublecolor=sgray](3,15)(9,15)
\psline[linearc=2\linearc,doublecolor=sgray](3,9)(9,9)
\psline[linearc=2\linearc,doublecolor=sgray](3,3)(9,3)
\threevertex{4}{12}{90}
\threevertex{4}{0}{-90}
\psset{doubleline=true}
\psbezier(4,13)(4,14)(3,14)(3,13)
\psbezier(3,13)(3,10)(4,10)(4,7)
\fourvertex{4}{6}{0}
\psset{doubleline=true}
\psbezier(4,5)(4,2)(3,2)(3,-1)
\psbezier(4,-1)(4,-2)(3,-2)(3,-1)
\psline(5,6)(6,6)
\fourvertex{7}{6}{0}
\psset{doubleline=true}
\psline(8,6)(9,6)
%\psbezier(7,7)(7,8)(9,8)(9,6)
%\psline(9,7)(9,5)
%\psbezier(7,5)(7,4)(9,4)(9,6)
\psline(5,12)(9,12)
\psline(5,0)(9,0)
\psline[linearc=2\linearc](7,7)(7,18)(15,18)(15,-3)(7,-3)(7,5)
\end{pspicture}}%
%\begin{pspicture}(-1,-4)(13,16)%\showgrid
%%\psframe(-1,-4)(13,13)
%\ulinsert[sgray]{0}{15}
%\olvertex{0}{12}
%\olvertex[sgray]{0}{9}
%\olvertex{0}{6}
%\olvertex[sgray]{0}{3}
%\dlinsert{0}{0}
%\drinsert{12}{0}
%\orvertex[sgray]{12}{3}
%\orvertex{12}{6}
%\orvertex[sgray]{12}{9}
%\orvertex{12}{12}
%\urinsert[sgray]{12}{15}
%\psset{doubleline=true}
%\psline[linearc=2\linearc,doublecolor=sgray](3,15)(9,15)
%\psline[linearc=2\linearc,doublecolor=sgray](3,9)(9,9)
%\psline[linearc=2\linearc,doublecolor=sgray](3,3)(9,3)
%\threevertex{6}{12}{90}
%\threevertex{6}{0}{-90}
%\psset{doubleline=true}
%\psline(3,12)(5,12)
%\psbezier(4,13)(4,14.25)(6,14.25)(6,13)
%\psbezier(4,13)(4,10)(4,10)(4,7)
%\fourvertex{4}{6}{0}
%\psset{doubleline=true}
%\psline(3,0)(5,0)
%\psbezier(4,5)(4,2)(4,2)(4,-1)
%\psbezier(4,-1)(4,-2.25)(6,-2.25)(6,-1)
%\psline(5,6)(6,6)
%\fourvertex{7}{6}{0}
%\psset{doubleline=true}
%\psline(8,6)(9,6)
%\psbezier(7,7)(7,8)(9,8)(9,6)
%%\psline(9,7)(9,5)
%\psbezier(7,5)(7,4)(9,4)(9,6)
%\psline(7,12)(9,12)
%\psline(7,0)(9,0)
%\end{pspicture}}%
\end{center}
\caption{
Some examples of higher genus diagrams with $R=2$
\subref{genus2nonwrap}-\subref{genus1wrap2} and $R=3$ \subref{genus2wrap}.
Their spectator connections are depicted in gray.
A genus $H=1$
  non-wrapping diagram \subref{genus2nonwrap}.
A genus $H=1$ wrapping diagram \subref{genus1wrap1} where both spectator
  connections increase the genus $H$ by one to $H=3$. 
A genus $H=1$ wrapping diagram \subref{genus1wrap2} where only the upper
spectator connection increases the genus $H$ by one to $H=2$.
No further handle has to be added for the downer spectator line to resolve its
crossings, since this is done by the handle that resolves the crossing
between the two field lines.
A genus $H=2$ wrapping diagram \subref{genus2wrap}, in which a non-planar
self energy contribution of an internal line is responsible for the
genus reduction. The two handles for the other field lines resolve the
crossings of the line that runs along the wrapping path.}
\label{genus12diag}
\end{figure}
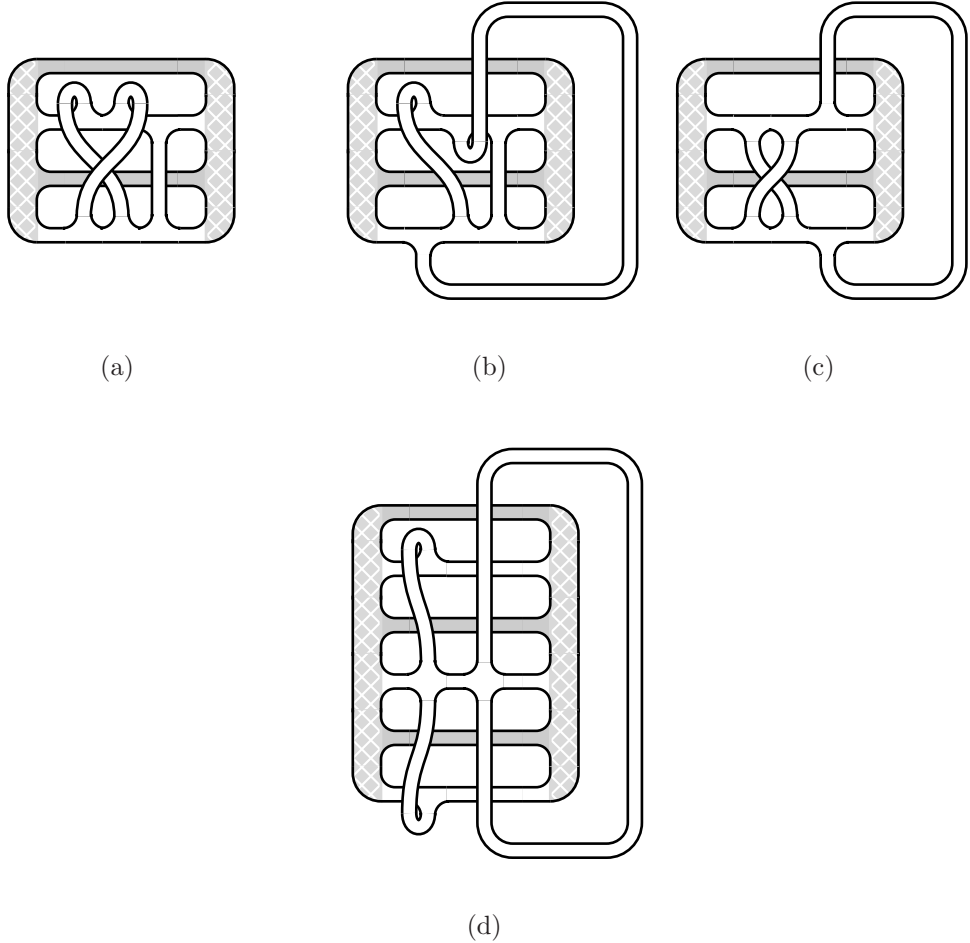

Restricting to genus $h\le1$ contributions to the Green function $V_{2R}$,
one finds from the considerations in Subsection \ref{spectatorfields} that a
necessary and sufficient condition for a diagram to be wrapping [non-wrapping] 
is that there exists no [at least one] possibility 
to add a pair of spectator fields $\cpair[0.1cm]{\psi}{\hat\psi}$ to
the diagram such that the crossings of their connection line with the
other field lines do not require adding one handle. 
We should remark that in this statement the restriction to $h\le 1$ is
essential. For $h\ge2$ one can find
counter-examples to this statement, in which the handles 
allow one to add spectator lines to wrapping diagrams without changing
their genera. 
We have depicted some examples in Fig.\ \ref{genus12diag}.

Instead of adding one pair of spectator fields, 
one can directly put $R$ pairs of spectator fields
into the $R$ positions between the $\phi_{n_r}$ of each of the two operators  
$\O{e}{2R}$.
No crossings between spectator connections are allowed.
Keeping fixed the position of the $\psi$, there are $R$
possible cyclic permutations for the $\hat\psi$. This means that to 
each of the $R^2$ contributions to   
$\big(\O{1}{R},V_{2R},\O{2}{R}\big)$ that come from the cyclic
permutations of the operators $\O{e}{2R}$ in \eqref{codef}, 
one has to associate $R$ variants by cyclic permutations only of the
spectators \emph{within} $\O{2}{2R}$,
keeping \emph{fixed} the fields $\phi_{n_r}$ in its trace.

With the help of these $R$ variants of the spectator structure, one can now 
distinguish between wrapping and non-wrapping diagrams. 
In the case of a wrapping diagram, all
pairwise connections between the spectators in each of the $R$ variants
crosses at least one other field line. 
In case of a non-wrapping diagram, there is at least one
variant in which one spectator line obstructs the unoccupied 
wrapping path such that only up to $R-1$ spectator lines cross other lines.

%The practical advantage of identifying a wrapping diagram by observing the
%behavior of spectator lines becomes clear if one restricts 
%oneself to the case of planar wrapping diagrams. These are the genus $h=1$
%contributions to $V_{2R}$ that contribute to
%$\big(\O{1}{R},V_{2R},\O{2}{R}\big)^{(0)}$.
In the case of planar contributions
$\big(\O{1}{R},V_{2R},\O{2}{R}\big)^{(0)}$ to the $2$-point function, 
each connection of a pair
of spectators that crosses other lines increases the genus $H$ by one. 
It is important 
to stress that only in the planar case one necessarily has to add a handle for
each spectator line that crosses other lines. For higher genus diagrams this
correspondence breaks down as is shown by some examples in Fig.\
\ref{genus12diag}. One reason for this is that one handle can be used not only
to cross another line, but even to resolve an additional crossing between 
lines that run along this handle. For $H\ge1$ there seems to be no
universal way to identify the connections that are responsible for the
genus reduction.

The result from the above considerations is that a diagram $\Dw{K}{1}{2R}$ is 
a planar wrapping diagram if the genus of the corresponding diagram of the
$2$-point function $\big(\O{1}{2R},\Dw{K}{1}{2R},\O{2}{2R}\big)$ with $R$
spectator pairs is always increased by $R$, i.e. to $H=R$.
Here we have introduced the modified $2$-point function
$\big(\O{1}{2R},V_{2R},\O{2}{2R}\big)$, where the difference in the 
number of elementary fields at both $\O{e}{2R}$ and $V_{2R}$ is given by the 
number of spectator fields.
In case of genus $H\ge1$, this statement is sufficient for a diagram to
be wrapping, but it is not necessary, as argued above.
% with the example given in Fig.\ \ref{genus12diag}.

The classification via spectators is now used to project out the wrapping
contributions. For this purpose, consider the connected ($C=1$) genus $h=1$ 
contributions $\V{}{1}{2R}$ to $V_{2R}$. 
They generate planar wrapping contributions and 
genus $H=1$ non-wrapping contributions to the $2$-point function
$\big(\O{1}{R},V_{2R},\O{2}{R}\big)$.
We now add $R$ pairs of spectators in $R$ different ways as described above
and obtain the modified $2$-point function
$\big(\O{1}{2R},V_{2R},\O{2}{2R}\big)$.
We focus on one of its contributions generated by $\V{}{1}{2R}$. 
If all $R$ cyclic permutations of the $R$ spectator fields added to one of the 
operators, lets say $\hat\psi$ in $\O{2}{2R}$, lead to genus $H=R$
contributions, the contribution is a wrapping one. 
Applying this procedure in case of non-wrapping diagrams,
one finds contributions of genus $H=R-1,\dots R+1$.\\
\underline{$H=R-1$:} 
One spectator line obstructs the unoccupied wrapping path. 
Another spectator line does not cross the other field lines because of the 
single handle in the $h=1$ contributions.\\
\underline{$H=R$:} 
Either one spectator line obstructs the unoccupied wrapping
path 
and all other spectator lines cross other field lines, or the wrapping path 
remains unobstructed and one spectator line does not
cross the other field lines 
because of the single handle.\\ 
\underline{$H=R+1$:} 
No spectator line obstructs the wrapping path and all other
spectator lines cross other field lines.

One thus has the relation
\begin{equation}\label{wrapnonwrapspec2ptdecomp}
%\begin{aligned}
\big(\O{1}{2R},\V{}{1}{2R},\O{2}{2R}\big)=\sum_{\delta=-1}^1\big(\O{1}{2R},\Vnw{}{1}{2R},\O{2}{2R}\big)^{(R+\delta)}
%+\big(\O{1}{2R},\Vnw{}{1}{2R},\O{2}{2R}\big)^{(R+1)}\\
%\phantom{{}={}}
+\big(\O{1}{2R},\Vw{}{1}{2R},\O{2}{2R}\big)^{(R)}\pnt
%\end{aligned}
\end{equation}
for the $C=1$ contributions to $V_{2R}$.\footnote{For $C\ge2$, the sum
  includes $\delta\le-2$ and the wrapping contribution is absent.} 
The above equation is now evaluated as follows:
one first adds the $R$ spectator pairs to the operators $\O{e}{2R}$, only 
ensuring that their connections never cross each other. Then, for a given 
contribution to $V_{2R}$ one performs all contractions of the $\phi_{n_r}$,
using the rules \eqref{ffrulesshortnot}.

In this way, one obtains an expression in which the $R$ spectator pairs are
distributed among a number of $1\le T\le 2R$ traces.
These distributions encode a classification of the diagrams. 
To see this, we interpret  
each trace $\big(\psi_1\,\dots\,\psi_{s_t}\big)$ over $s_t$ spectator fields 
as a vertex with $s_t$ directed legs. The direction distinguishes legs
associated to $\psi$ from legs which represent $\hat\psi$. 
Each pairwise contraction is regarded as a propagator 
connecting an outgoing with an ingoing leg of these vertices. 
As usual, these diagrams can be classified by the 
number of vertices, in this case given by $T$, the number of connected pieces 
$C$, the number of loops $L$. 
The number of propagators is fixed to $P=R$ and the number of
external legs is $E=0$.
In this picture, the removal of a spectator pair corresponds to a removal 
of a propagator. The execution of a contraction between pairs either becomes 
the fusion of two vertices or the fission of a single vertex into two,
depending on whether the propagator contracts two legs at distinct vertices or
at only one vertex. 
The diagrams that are generated from wrapping contributions are now identified
as follows: Each removal of a pair $\psi$, $\hat\psi$ must
decrease the genus $H$ by
one and hence lead to an additional factor $N^2$. Thus, 
each removal of a spectator pair must increase the trace number $T$ by
one, compared to the case where this pair is kept and finally contracted.
Each trace at the very 
end contributes a factor $\big(\unitmatrix\big)=N$.
Another $N$-dependent factor comes from changing
the normalization of the operators \eqref{codef} that is required
  when the number of fields is changed. 
Furthermore, the above given issues of the spectator diagrams should
remain true if all cyclic permutations of the $\hat\psi$ are taken into
account. Hence, for the spectator diagrams of planar wrapping 
contributions a necessary condition is that no contractions must occur 
within a single trace. 
Otherwise, the removal of this pair of spectators would not change the genus.
Instead of avoiding the fusion of two traces the fission of that
  trace would be inhibited. 
Since such a configuration must not occur in any of the cyclic 
permutations, the spectators within one trace must only be of one type, 
either $\psi$ or $\hat\psi$. Furthermore, the increasing of the trace number
by removing one spectator pair must be independent of the contraction or
removal of other spectators.
This forbids that the connections form any closed 
loops since, after contracting some of them, one always would end up in a 
contraction of one spectator pair within a single trace. 
The spectator diagrams of wrapping contributions 
are therefore always tree-level and consist of $1\le C\le R$ separate pieces.
Any contraction of a pair $\psi$, $\hat\psi$ is an application of the fusion 
rule \eqref{ffrulesshortnot} such that, after having contracted 
all spectator pairs, $C$ separate
traces $\big(\unitmatrix\big)$ remain, leading to a factor $N^C$. 
On the other hand, removing all pairs $\psi$, $\hat\psi$, one is left with $T$
traces $\big(\unitmatrix\big)$, contributing $N^T$. A factor $N^R$ has to be
considered for the change in the normalization of the operators. 
According to the last equation in \eqref{diageq}, in the special case of 
tree-level diagrams one has the relation $R=T-C$,
and hence concludes:\\
Each tree-level spectator diagram that is built out of $C$ connected pieces 
contributes a factor $N^C$ if the spectators are kept and it contributes a
factor $N^{2R+C}$ if the spectators are removed. 

In our case $C$ has to be maximal, i.e. $C=R$. One can find several arguments
for this statement. First of all, removing all the spectator pairs, one
obtains a planar diagram, and planar diagrams contribute to the leading power
in the $\frac{1}{N}$ expansion. This means, if $C$ were not maximal for the 
planar wrapping diagrams, their contribution would be
subdominant.
Secondly, one can regard a diagram $\D{K}{h}{2R}$ as an effective vertex which 
is proportional to $\sqrt{\lambda}^K$. 
The dependence on $N$ one should associate with such a vertex can be
interpreted as the $N$ power which has to be absorbed into an effective
't Hooft coupling that is later set to $1$. 
For a planar vertex with $2R$ external legs one has to absorb a factor 
$\sqrt{N}^{2R-2}$, i.e. the $N$ dependence is given by its reciprocal value
$N^{1-R}$. 
When this effective vertex contains $h$ handles, and it is 
contracted with a planar effective vertex in a planar way, the number 
of traces is reduced by $h$. This means, compared to the case of a
planar diagram which contains two planar vertices, the power in $N$ found
from the traces is reduced by $h$.
Therefore, to remove a factor $N^2$ for each handle in the final
result, one has to change the $N$ dependent normalization of an
effective vertex of genus $h$ to $N^{1-R-h}$.
That means, in case of planar wrapping
contributions which have $h=1$, including the traces over all spectator pairs, 
and the normalization of the two operators $\O{e}{2R}$ taken from
\eqref{codef}, the total $N$ dependence is $N^{2+C-3R}$. This factor
should be equal to $N^{2-2H}$ with $H=R$ (see
\eqref{wrapnonwrapspec2ptdecomp}), such that one deduces
$C=R$.\footnote{This argument is only true if the contractions of the
$\phi_{n_r}$ never generate powers in $N$. According to
\eqref{Tcompleterel}, this would happen if two contractible
fields $\phi_{n_r}$ become direct neighbours within one trace without further
fields. That this cannot happen here is guaranteed by the spectators that 
separate all $\phi_{n_r}$ from each other.}

From the above considerations it follows that one can uniquely 
identify the wrapping diagrams that contribute to the planar $2$-point
function by finding the spectator structure
\begin{equation}\label{planarwrapspecdiag}
\big(\rnode{1}{\psi}\big)\big(\rnode{2}{\psi}\big)\dots\big(\rnode{3}{\psi}\big)\big(\rnode{4}{\hat\psi}\big)\dots\big(\rnode{5}{\hat\psi}\big)\big(\rnode{6}{\hat\psi}\big)
\ncbar[linewidth=0.5pt,nodesep=1pt,angle=-90,arm=0.15cm]{3}{4}
\ncbar[linewidth=0.5pt,nodesep=1pt,angle=-90,arm=0.375cm]{2}{5}
\ncbar[linewidth=0.5pt,nodesep=1pt,angle=-90,arm=0.6cm]{1}{6}
\settodepth{\armlen}{$\psi$}
\addtolength{\armlen}{0.6cm}
\addtolength{\armlen}{1.5pt}
\rule[-\armlen]{0pt}{\armlen}
=
\big(\rnode{1}{\psi}\big)\big(\rnode{2}{\hat\psi}\big)\big(\rnode{3}{\psi}\big)\big(\rnode{4}{\hat\psi}\big)\dots\big(\rnode{5}{\psi}\big)\big(\rnode{6}{\hat\psi}\big)
\ncbar[linewidth=0.5pt,nodesep=1pt,angle=-90,arm=0.15cm]{1}{2}
\ncbar[linewidth=0.5pt,nodesep=1pt,angle=-90,arm=0.15cm]{3}{4}
\ncbar[linewidth=0.5pt,nodesep=1pt,angle=-90,arm=0.15cm]{5}{6}
\settodepth{\armlen}{$\psi$}
\addtolength{\armlen}{0.15cm}
\addtolength{\armlen}{1.5pt}
\rule[-\armlen]{0pt}{\armlen}\pnt
\end{equation}
The issue that all traces contain only one spectator field enables one
in principle to directly project out the 
planar wrapping contributions generated by $\V{}{1}{2R}$. 
One has to find an appropriate matrix for $\psi$ and $\hat\psi$ such
that traces which include more than one spectator field vanish.
Such traces are always included in the remaining
genus $h=1$ non-wrapping contributions.
This issue is discussed in Section \ref{candpsi}.

\section{Planar contributions to the $2$-point function}
\label{planarcontrib}
\subsection{Non-interacting case}
\label{noninteractingcase}
The simplest contribution to the $2$-point function 
$\big(\O{1}{R},V_{2R},\O{2}{R}\big)$ is given by a pairwise planar ($h=0$)
connection of all the legs of the two operators without any interactions. 
The corresponding contribution to $V_{2R}$ 
is given by the diagram $\D{0}{0}{2R}$ which is a direct product of free 
propagators that connect the fields with label $r$ and $2R-r+1$,
$r=1,\dots R$.
The color part of this contribution, in case of a $U(N)$ gauge group, 
is given by
\begin{equation}\label{planarfreeV}
\big(\O{1}{R},\D{0}{0}{2R},\O{2}{R}\big)\to R^2\frac{1}{N^{R-2}}
\big(\rnode{1}{a_1}\,\dots\,
\rnode{2}{a_R}\big)\big(\rnode{3}{a_{R+1}}\,\dots\,
\rnode{4}{a_{2R}}\big)=R^2N\big(\unitmatrix\big)=R^2N^2\col
\ncbar[linewidth=0.5pt,nodesep=1pt,angle=-90,arm=0.15cm]{2}{3}
\ncbar[linewidth=0.5pt,nodesep=1pt,angle=-90,arm=0.6cm]{1}{4}
\settodepth{\armlen}{$a_{R+1}$}
\addtolength{\armlen}{0.6cm}
\addtolength{\armlen}{1.5pt}
\rule[-\armlen]{0pt}{\armlen}
\end{equation}
where the factor $R^2$ stems from the $R^2$ pairs of cyclic permutations 
of the fields in each of the two operators $\O{1}{R}$. The 
second equality in \eqref{planarfreeV} follows after applying the fusion rules
for the traces \eqref{ffrulesshortnot} and the identity \eqref{Tcompleterel} 
with $a_0=0$.

The $N$ dependent prefactor is the square of the normalization of the 
operators in \eqref{codef}, and it is chosen such that the above result is
of the order $N^2$. A genus $H$ contribution to
$\big(\O{1}{R},V_{2R},\O{2}{R}\big)$ will then be of the order 
$N^{2-2H}$.

\subsection{Generic planar connected contributions to $V_{2R}$}

\begin{figure}
\begin{center}
\begin{pspicture}(0,-1)(12,10)%\showgrid
%\psframe(0,-1)(12,10)
\uinex{2}{9}
\iinex{2}{6}
%\iinex{3}{3}
\dinex{2}{0}
\doutex{10}{0}
%\ioutex{9}{3}
\ioutex{10}{6}
\uoutex{10}{9}
\setlength{\ya}{9\unit}
\addtolength{\ya}{0.5\dlinewidth}
\psline(3.5,\ya)(8.5,\ya)
\setlength{\yb}{0\unit}
\addtolength{\yb}{-0.5\dlinewidth}
\psline(3.5,\yb)(8.5,\yb)
\psline[linestyle=dotted](3.5,4.5)(3.5,1.5)
\psline[linestyle=dotted](8.5,4.5)(8.5,1.5)
\setlength{\xa}{3.5\unit}
\addtolength{\xa}{\dlinewidth}
\setlength{\xb}{8.5\unit}
\addtolength{\xb}{-\dlinewidth}
\setlength{\ya}{9\unit}
\addtolength{\ya}{-0.5\dlinewidth}
\setlength{\yb}{0\unit}
\addtolength{\yb}{0.5\dlinewidth}
\pscustom[linecolor=gray,fillstyle=solid,fillcolor=gray]{%
\psline[liftpen=1,linearc=\linearc](\xa,4.5)(\xa,\ya)(\xb,\ya)(\xb,4.5)}
\pscustom[linecolor=gray,fillstyle=solid,fillcolor=gray]{%
\psline[liftpen=1,linearc=\linearc](\xa,1.5)(\xa,\yb)(\xb,\yb)(\xb,1.5)}
\psline[linestyle=dotted,linecolor=gray](\xa,4.5)(\xa,1.5)
\psline[linestyle=dotted,linecolor=gray](\xb,4.5)(\xb,1.5)
%\psframe[framearc=0.5](\xa,\ya)(\xb,\yb)
\end{pspicture}
\end{center}
\caption{A generic planar connected contribution to $V_{2R}$. The picture
  shows that a planar contribution contracts the two 
    index lines of one external line with one of the index lines of the two
    neighboured external lines. This corresponds to a contraction of
    the indices of the representation matrices. The gray-filled structure in
    the middle represents any planar diagram. It has no influence on the index
    lines that belong to the external legs. Thus all planar contributions
    consists of a single trace over the representation matrices.}
\label{planarVdiag}
\end{figure}
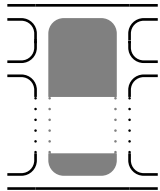

Consider the planar connected contributions to $V_{2R}$ denoted as
$\V{}{0}{2R}$.  Their general form is shown in Fig.\ \ref{planarVdiag}. 

From the fact that
in the planar case lines must not cross in the corresponding ribbon graphs, 
it is obvious that they can always be represented as an effective
vertex of the form 
\begin{equation}\label{planarV}
\V{}{0}{2R}
%=\f{0}{2R}\big(\sqrt\lambda\big)\frac{1}{N^{R-1}}\tr(\mathfrak{t}^{a_{2R}}\dots\mathfrak{t}^{a_1})
=\f{0}{2R}\big(\sqrt\lambda\big)\frac{1}{N^{R-1}}\big(a_{2R}\,\dots\, a_1\big)\col
\end{equation}
where $\f{0}{2R}\big(\sqrt\lambda\big)$ captures the individual properties of the
diagrams, i.e. their coupling dependence and the contributions from the loop
integrals. The $N$ dependence has been fixed such that 
any planar diagram including this vertex is of the order $N^2$.
%One can obtain this factor by introducing $R$ propagators that simply contract
%the legs of the vertex as $\cpair[0.1cm]{a_r}{a_{2R-r+1}}$, $r=1,\dots R$.

%\subsection{Spectator diagram}
Including $R$ spectator pairs $\cpair[0.1cm]{\psi}{\hat\psi}$,
the generic planar contribution \eqref{planarV} to the $2$-point
function is given by 
\begin{equation}\label{planarspecdiag}
\begin{aligned}
&\big(\O{1}{2R},\V{}{0}{2R},\O{2}{2R}\big)\\
&\qquad\propto\big(\rnode{a1}{a_1}\,\rnode{1}{\psi}\,\rnode{a2}{a_2}\,\dots\,\rnode{Rm1}{\psi}\,\rnode{aR}{a_R}\,\rnode{R}{\psi}\big)\big(a_{R}\,\dots\,a_{1}\,a_{2R}\,\dots\,a_{R+1}\big)\big(\rnode{Rp1}{\hat\psi}\,\rnode{aRp1}{a_{R+1}}\,\rnode{Rp2}{\hat\psi}\,\dots\,\rnode{a2Rm1}{a_{2R-1}}\,\rnode{2R}{\hat\psi}\,\rnode{a2R}{a_{2R}}\big)
%+\text{cycl.\ perm.\ $\hat\psi$}
\ncbar[linewidth=0.5pt,nodesep=1pt,angle=-90,arm=0.15cm]{R}{Rp1}
\ncbar[linewidth=0.5pt,nodesep=1pt,angle=-90,arm=0.375cm]{Rm1}{Rp2}
\ncbar[linewidth=0.5pt,nodesep=1pt,angle=-90,arm=0.6cm]{1}{2R}
\settodepth{\armlen}{$a_{R+1}$}
\addtolength{\armlen}{0.6cm}
\addtolength{\armlen}{1.5pt}
\rule[-\armlen]{0pt}{\armlen}\\
&\qquad\phantom{{}={}}+\text{cycl.\ perm.\ $\hat\psi$}
\settodepth{\armlen}{$a_{R+1}$}
\addtolength{\armlen}{0.6cm}
\addtolength{\armlen}{1.5pt}
\rule[-\armlen]{0pt}{\armlen}\\
&\qquad=\big(\rnode{1}{\psi}\big)\dots\big(\rnode{Rm1}{\psi}\big)\big(\rnode{R}{\psi}\,\rnode{Rp1}{\hat\psi}\big)\big(\rnode{Rp2}{\hat\psi}\big)\dots\big(\rnode{2R}{\hat\psi}\big)+\text{cycl.\
  perm.\ $\hat\psi$}\col
\ncbar[linewidth=0.5pt,nodesep=1pt,angle=-90,arm=0.15cm]{R}{Rp1}
\ncbar[linewidth=0.5pt,nodesep=1pt,angle=-90,arm=0.375cm]{Rm1}{Rp2}
\ncbar[linewidth=0.5pt,nodesep=1pt,angle=-90,arm=0.6cm]{1}{2R}
\settodepth{\armlen}{$\psi$}
\addtolength{\armlen}{0.6cm}
\addtolength{\armlen}{1.5pt}
\rule[-\armlen]{0pt}{\armlen}
\end{aligned}
\end{equation}
where in the first equality we have used the invariance of the trace under
cyclic permutations for a reordering of \eqref{planarV}.
The contribution given explicitly in the second equality of 
\eqref{planarspecdiag}
is the one that identifies the diagram as non-wrapping, since the removal of
the spectator pair in the single trace does not increase the power in $N$ and 
hence does not decrease the genus of the diagram. The connection line
of this spectator pair obstructs the wrapping path.
The other contributions to \eqref{planarspecdiag} are of the form
\begin{equation}\label{planarspecdiag2}
\big(\rnode{1}{\psi}\big)\dots\big(\rnode{Rm1}{\psi}\big)\big(\rnode{R}{\psi}\,\rnode{Rp1}{\hat\psi}\big)\big(\rnode{Rp2}{\hat\psi}\big)\dots\big(\rnode{2R}{\hat\psi}\big)
\ncbar[linewidth=0.5pt,nodesep=1pt,angle=-90,arm=0.15cm]{R}{2R}
\ncbar[linewidth=0.5pt,nodesep=1pt,angle=-90,arm=0.375cm]{Rm1}{Rp1}
\ncbar[linewidth=0.5pt,nodesep=1pt,angle=-90,arm=0.6cm]{1}{Rp2}
\settodepth{\armlen}{$\psi$}
\addtolength{\armlen}{0.6cm}
\addtolength{\armlen}{1.5pt}
\rule[-\armlen]{0pt}{\armlen}
\end{equation}
In contrast to first contribution in the second line of
\eqref{planarspecdiag}, the removal of any spectator pair increases the power
in $N$ by two and hence reduces the genus of the corresponding diagram by
one. In these contributions, no spectator line obstructs the wrapping path.

\subsection{Generic planar wrapping contributions to $V_{2R}$}

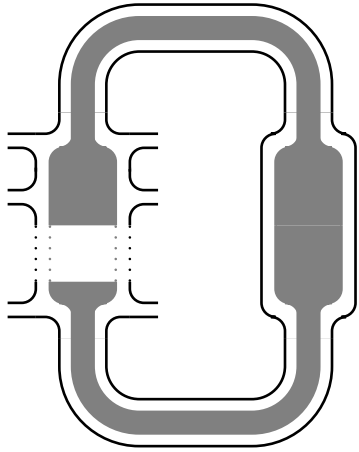
\begin{figure}
\begin{center}
\begin{pspicture}(0,-9)(22.5,18)%\showgrid
%\psframe(0,-9)(22.5,18)
\uinex{2}{9}
\iinex{2}{6}
%\iinex{3}{3}
\dinex{2}{0}
\doutex{10}{0}
%\ioutex{9}{3}
\ioutex{10}{6}
\uoutex{10}{9}
\psline[linestyle=dotted](3.5,4.5)(3.5,1.5)
\psline[linestyle=dotted](8.5,4.5)(8.5,1.5)
\setlength{\xa}{4.75\unit}
%\addtolength{\xa}{0.125\dlinewidth}
\setlength{\xb}{7.25\unit}
%\addtolength{\xb}{-0.125\dlinewidth}
\setlength{\xc}{3.5\unit}
\addtolength{\xc}{\dlinewidth}
\setlength{\xd}{8.5\unit}
\addtolength{\xd}{-\dlinewidth}
\setlength{\ya}{9\unit}
\addtolength{\ya}{-0.5\dlinewidth}
\setlength{\yb}{0\unit}
\addtolength{\yb}{0.5\dlinewidth}
\setlength{\yc}{\ya}
\addtolength{\yc}{-0.5\dlinewidth}
\setlength{\yd}{\yb}
\addtolength{\yd}{0.5\dlinewidth}
\newlength{\wsep}
\setlength{\wsep}{\xb}
\addtolength{\wsep}{-\xa}
\addtolength{\wsep}{-\linew}
\newlength{\wwidth}
\setlength{\wwidth}{\xb}
\addtolength{\wwidth}{-\xa}
\addtolength{\wwidth}{-2\dlinewidth}
\newlength{\yup}
\setlength{\yup}{9\unit}
\addtolength{\yup}{0.5\dlinewidth}
\psline[linearc=\linearc](3.5,\yup)(\xa,\yup)(\xa,10.5)
\psline[linearc=\linearc](8.5,\yup)(\xb,\yup)(\xb,10.5)
\newlength{\ydown}
\setlength{\ydown}{0\unit}
\addtolength{\ydown}{-0.5\dlinewidth}
\newlength{\xileft}
\setlength{\xileft}{\xa}
\addtolength{\xileft}{\dlinewidth}
\newlength{\xiright}
\setlength{\xiright}{\xb}
\addtolength{\xiright}{-\dlinewidth}
\newlength{\yiup}
\setlength{\yiup}{9\unit}
\addtolength{\yiup}{-0.5\dlinewidth}
\newlength{\yidown}
\setlength{\yidown}{0\unit}
\addtolength{\yidown}{0.5\dlinewidth}
\psline[linearc=\linearc](3.5,\ydown)(\xa,\ydown)(\xa,-1.5)
\psline[linearc=\linearc](8.5,\ydown)(\xb,\ydown)(\xb,-1.5)
\psline[linearc=4\linearc,doubleline=true,doublesep=\wsep](6,10.5)(6,15)(18,15)(18,10.5)
\psset{origin={-12,0}}
\psline[liftpen=1,linearc=\linearc](4.125,\yup)(\xa,\yup)(\xa,10.5)
\psline[linearc=\linearc](7.875,\yup)(\xb,\yup)(\xb,10.5)
\psline[liftpen=1,linearc=\linearc](4.125,\ydown)(\xa,\ydown)(\xa,-1.5)
\psline[linearc=\linearc](7.875,\ydown)(\xb,\ydown)(\xb,-1.5)
\psline[liftpen=1,linearc=\linearc](4.125,\yup)(3.5,\yup)(3.5,\ydown)(4.125,\ydown)
\psline[linearc=\linearc](7.875,\yup)(8.5,\yup)(8.5,\ydown)(7.875,\ydown)
\psset{origin={0,0}}
\psline[liftpen=1,linearc=4\linearc,doubleline=true,doublesep=\wsep]
(18,-1.5)(18,-6)(6,-6)(6,-1.5)
\pscustom[linecolor=gray,fillstyle=solid,fillcolor=gray]{%
\psline[liftpen=1,linearc=\linearc](\xc,4.5)(\xc,\ya)(\xileft,\ya)(\xileft,10.5)
\psline[liftpen=1,linearc=\linearc](\xiright,10.5)(\xiright,\ya)(\xd,\ya)(\xd,4.5)}
\pscustom[linecolor=gray,fillstyle=solid,fillcolor=gray]{%
\psline[liftpen=1,linearc=\linearc](\xc,1.5)(\xc,\yb)(\xileft,\yb)(\xileft,-1.5)\psline[liftpen=1,linearc=\linearc](\xiright,-1.5)(\xiright,\yb)(\xd,\yb)(\xd,1.5)}
\psline[linestyle=dotted,linecolor=gray](\xc,4.5)(\xc,1.5)
\psline[linestyle=dotted,linecolor=gray](\xd,4.5)(\xd,1.5)
\psline[liftpen=1,linearc=4\linearc,doubleline=true,doublesep=\wwidth,doublecolor=gray,linecolor=gray](6,\ya)(6,15)(18,15)(18,\ya)
\psset{origin={-12,0}}
\pscustom[linecolor=gray,fillstyle=solid,fillcolor=gray]{%
\psline[liftpen=1,linearc=\linearc](\xc,4.5)(\xc,\ya)(\xileft,\ya)(\xileft,10.5)
\psline[liftpen=1,linearc=\linearc](\xiright,10.5)(\xiright,\ya)(\xd,\ya)(\xd,4.5)}
\pscustom[linecolor=gray,fillstyle=solid,fillcolor=gray]{%
\psline[liftpen=1,linearc=\linearc](\xc,4.5)(\xc,\yb)(\xileft,\yb)(\xileft,-1.5)\psline[liftpen=1,linearc=\linearc](\xiright,-1.5)(\xiright,\yb)(\xd,\yb)(\xd,4.5)}
\psset{origin={0,0}}
\psline[liftpen=1,linearc=4\linearc,doubleline=true,doublesep=\wwidth,doublecolor=gray,linecolor=gray](18,\yb)(18,-6)(6,-6)(6,\yb)
%(6,15)(18,15)(18,10.5)%(18,-6)(6,-6)(6,-1.5)
%\psline[liftpen=0,linearc=3\linearc,linewidth=\wwidth,linecolor=gray](6,4.5)
%(6,15)(15,15)(15,-6)(6,-6)(6,1.5)
%\psline[linestyle=dotted,linecolor=gray](\xa,4.5)(\xa,1.5)
%\psline[linestyle=dotted,linecolor=gray](\xb,4.5)(\xb,1.5)
%\psframe[framearc=0.5](\xa,\ya)(\xb,\yb)
\end{pspicture}
\end{center}
\caption{A generic planar wrapping contribution to $V_{2R}$. 
The occupation of the wrapping path
separates the ingoing and outgoing fields from each other and breaks the
single trace of ordinary planar diagrams in Fig.\ \ref{planarVdiag} into two 
traces. The gray-filled structure in the 
middle represents any planar diagram. It has no influence on the lines
that belong to the external legs. Thus all planar wrapping contributions 
consists of a product of two separate traces over the representation matrices 
for the ingoing and outgoing legs.}
\label{planarVwrapdiag}
\end{figure}

Consider the planar wrapping contributions to $V_{2R}$ denoted by
$\Vw{}{1}{2R}$. In a genus expansion of the $2$-point function 
$\big(\O{1}{R'},V_{2R},\O{2}{R'}\big)$ they contribute at planar level
if $R'=R$, but all their contributions become genus $H=1$ for
$R'>R$. The most general planar wrapping contribution is obtained by 
taking the most general planar diagram and then filling the wrapping path 
with the most general planar structure. One then arrives at the 
ribbon graph shown in Fig.\ \ref{planarVwrapdiag}. 

From the restriction that
lines must not cross in this graph, one can read off the general effective
vertex for a generic planar wrapping diagram.
It is given by
\begin{equation}\label{planarVwrap}
\begin{aligned}
\Vw{}{1}{2R}
%&=\fw{1}{2R}\big(\sqrt{\lambda}\big)\frac{1}{N^R}\tr(\mathfrak{t}^{a_R}\dots\mathfrak{t}^{a_1})\tr(\mathfrak{t}^{a_{2R}}\dots\mathfrak{t}^{a_{R+1}})\\
&=\fw{1}{2R}\big(\sqrt{\lambda}\big)\frac{1}{N^R}\big(a_R\,\dots\,a_1\big)\big(a_{2R}\,\dots\,a_{R+1}\big)\col
\end{aligned}
\end{equation}
where $\fw{1}{2R}\big(\sqrt\lambda\big)$ captures the individual properties of
the wrapping diagrams, i.e. their coupling dependence and the contributions
from the loop integrals. The $N$ dependence has been fixed such that 
any diagram with two operators $\O{e}{2R'}$, $R'>R$ including this vertex is 
of the order $N^0$.
%One can obtain this factor by introducing $R$ propagators that contract
%the ingoing legs of the vertex with its outgoing legs such that these
%connections do not cross each other, e.g. 
%$\cpair[0.1cm]{a_r}{a_{2R-r+1}}$, $r=1,\dots R$. The result has to be
%proportional to $N^0$.
The vertex \eqref{planarVwrap} is obviously invariant under separate
cyclic permutations of the ingoing and outgoing legs. This is even
obvious in Fig.\ \ref{planarVwrapdiag}, since these cyclic permutations do
not change the structure represented by the gray-filled region. 
The symmetry found for wrapping diagrams is 
consistent with the observation, that the wrapping handle is
sufficient to resolve all those crossings within the ingoing and outgoing
external legs of a given diagram $\D{K}{1}{2R}$ that can be traced
back to cyclic permutations. This can be easily understood, looking at
Fig.\ \ref{Vinfty}, and adding the wrapping handle that connects the
regions above and below the vertex at $\infty$.  

%\subsection{Spectator diagram}
Including $R$ spectator pairs $\cpair[0.1cm]{\psi}{\hat\psi}$, the $2$-point
function $\big(\O{1}{2R},\Vw{}{1}{2R},\O{2}{2R}\big)$ of the generic planar
wrapping contribution \eqref{planarVwrap} is given by 
\begin{equation}\label{planarwrapspeccontrib}
\begin{aligned}
&\big(\O{1}{2R},\Vw{}{1}{2R},\O{2}{2R}\big)\\
&\quad\propto R\,\big(\rnode{a1}{a_1}\,\rnode{1}{\psi}\,\rnode{a2}{a_2}\,\dots\,\rnode{Rm1}{\psi}\,\rnode{aR}{a_R}\,\rnode{R}{\psi}\big)\big(a_{R}\,\dots\,a_{1}\big)\big(a_{2R}\,\dots\,a_{R+1}\big)\big(\rnode{Rp1}{\hat\psi}\,\rnode{aRp1}{a_{R+1}}\,\rnode{Rp2}{\hat\psi}\,\dots\,\rnode{a2Rm1}{a_{2R-1}}\,\rnode{2R}{\hat\psi}\,\rnode{a2R}{a_{2R}}\big)\\
%+\text{cycl.\ perm.\ $\hat\psi$}
\ncbar[linewidth=0.5pt,nodesep=1pt,angle=-90,arm=0.15cm]{R}{Rp1}
\ncbar[linewidth=0.5pt,nodesep=1pt,angle=-90,arm=0.375cm]{Rm1}{Rp2}
\ncbar[linewidth=0.5pt,nodesep=1pt,angle=-90,arm=0.6cm]{1}{2R}
\settodepth{\armlen}{$a_{R+1}$}
\addtolength{\armlen}{0.6cm}
\addtolength{\armlen}{1.5pt}
\rule[-\armlen]{0pt}{\armlen}\\
%&\qquad\phantom{{}={}}+\text{cycl.\ perm.\ $\hat\psi$}
%\settodepth{\armlen}{$a_{R+1}$}
%\addtolength{\armlen}{0.6cm}
%\addtolength{\armlen}{1.5pt}
%\rule[-\armlen]{0pt}{\armlen}\\
&\quad=R\,\big(\rnode{1}{\psi}\big)\dots\big(\rnode{Rm1}{\psi}\big)\big(\rnode{R}{\psi}\big)\big(\rnode{Rp1}{\hat\psi}\big)\big(\rnode{Rp2}{\hat\psi}\big)\dots\big(\rnode{2R}{\hat\psi}\big)
%+\text{cycl.\ perm.\ $\hat\psi$}\col
\ncbar[linewidth=0.5pt,nodesep=1pt,angle=-90,arm=0.15cm]{R}{Rp1}
\ncbar[linewidth=0.5pt,nodesep=1pt,angle=-90,arm=0.375cm]{Rm1}{Rp2}
\ncbar[linewidth=0.5pt,nodesep=1pt,angle=-90,arm=0.6cm]{1}{2R}
\settodepth{\armlen}{$\psi$}
\addtolength{\armlen}{0.6cm}
\addtolength{\armlen}{1.5pt}
\rule[-\armlen]{0pt}{\armlen}
\end{aligned}
\end{equation}
where the factor $R$ in the first equality is generated by the cyclic
permutations of the $\hat\psi$ that in this case do not lead to different
contributions. 
The final result of \eqref{planarwrapspecdiag}
is the one that identifies the diagram as planar wrapping, since each removal
of a spectator pair increases the number of traces and hence the power in $N$
by one. With the additional factor of $N$ from the change in the normalization
this leads to a factor $N^2$, showing the decreasing of the
genus $H$ of the diagram by one. 

\subsection{Generic planar wrapping contributions to $V_{2R}$ including flavor}
\label{genplanarwrapflavor}
 We extend the effective vertex \eqref{planarVwrap} of planar wrapping
 interactions by considering theories which include flavor
 degrees of freedom such that the flavor structure is described by
 Kronecker $\delta$s. In the subsector of composite operators which
 contain no trace terms in their flavor indices\footnote{This is true
 for instance in the $SU(2)$ subsector of $\mathcal{N}=4$ SYM,
 consisting of two scalars $X=\frac{1}{\sqrt{2}}(\phi_1+i\phi_2)$,
 $Y=\frac{1}{\sqrt{2}}(\phi_3+i\phi_4)$.} one can write down a rather
 simple expression for this vertex.
%the most general vertex describing the planar
% wrapping contributions to the $2$-point function
% $\big(\O{1}{R},V_{2R},\O{2}{R}\big)$. 
It is given by 
\begin{equation}
\begin{aligned}\label{planarVwrapflavor}
\Vw{}{1}{2R}
&=\frac{1}{N^R}\big(a_R\,\dots\,a_1\big)\big(a_{2R}\,\dots\,a_{R+1}\big)\\
&\phantom{={}}
\times\sum_{\pi\in\frac{S_R}{C_R}}\frac{\f{1}{\pi,2R}\big(\sqrt\lambda\big)}{R}\sum_{\omega\in C_R}\delta^{m_1}_{m_{\omega(\pi(2R))}}\delta^{m_2}_{m_{\omega(\pi(2R-1))}}\dots\delta^{m_R}_{m_{\omega(\pi(R+1))}}\col
\end{aligned}
\end{equation}
where $\f{1}{\pi,2R}\big(\sqrt\lambda\big)$ captures the individual properties
of the diagrams, and $m_r$ is the corresponding flavor index carried by the
leg with color index $a_r$. 
Some remarks should be made. First of all, the above expression is
valid to arbitrary high order in $\lambda$. Secondly, the vertex is chosen such
that its flavor dependence is invariant under 
separate cyclic permutations applied to the two pairs of $R$ ingoing and
outgoing legs. Since the color dependence respects the same symmetry, it is
a symmetry of the complete vertex. 
%This property seems to contradict the explicit results \eqref{11} and
%\eqref{14} for the wrapping terms in our toy example in Section \ref{toymodel}
%below. However, one has to take into account that for the construction of the
%$2$-point function \eqref{11} as well as \eqref{14} have to be contracted with
%all $R$ equivalent representations of each of the two operators $\O{1}{R}$
%obtained by applying the cyclic permutations. Therefore, one can choose 
%an effective vertex that respects the additional symmetry. 
%Its coefficient functions 
%$\f{0}{\pi,2R}\big(\sqrt\lambda\big)$ are given by the sum of all
%  coefficient functions of the flavor structures in the original
%  vertex that can be mapped to
%  each other by the above mentioned cyclic permutations.
%In the construction of the $2$-point function the effective vertex 
%\eqref{planarVwrapflavor} gives the same result as
%obtained by taking the original vertex. 

The choice of this effective vertex enables one to reduce the number of
coefficient functions from $R!$ in the original vertex to $(R-1)!$.
This reduction should be useful in practical calculations. 
Instead of explicitly computing the coefficient functions by summing up the 
diagrams of the perturbation expansion, one could try to determine the
coefficient functions by fitting them to available data. 
In this case, the chosen effective vertex should of course contain the minimum
number of coefficient functions. The choice of \eqref{planarVwrapflavor}
would be a first step. As a second step, one should try to use 
additional symmetries, unitarity, non-renormalization theorems, etc.
to further reduce the number
of independent coefficient functions.

\section{The toy model}
\label{toymodel}
Let us consider the following toy-example of a single scalar field in the 
adjoint representation of $U(N)$, with a standard
kinetic term and an interaction Lagrangian:
\begin{equation}
\label{1}
S_\text{int}[\phi]=\frac{g^2}{4!}\tr(\phi\phi\phi\phi)\pnt
\end{equation}
From here one can derive the interaction vertex as
\begin{equation}
\begin{aligned}
\label{2}
V^{a_1a_2a_3a_4}&=\frac{g^2}{3!}
\big[\big(a_1\,a_2\,a_3\,a_4\big)+\big(a_1\,a_2\,a_4\,a_3\big)
+\big(a_1\,a_3\,a_2\,a_4\big)\\
&\phantom{={}4g^2\big[}
+\big(a_2\,a_1\,a_3\,a_4\big)+\big(a_1\,a_4\,a_2\,a_3\big)+\big(a_1\,a_4\,a_3\,a_2\big)\big]\col  
\end{aligned}
\end{equation}
where we have used the abbreviation \eqref{traceabb} for the traces. 
Due to the cyclic invariance of the trace, only six inequivalent
permutations out of $4!=24$ possible permutations remain.

We compute the $\propto g^4$ contribution to the 
$2$-point function of the
composite operator $\O{}{2}=\tr(\phi^2)$ in perturbation theory.
As derived in Appendix \ref{countingrules}, this order ($K=4$)
is precisely the critical case for this operator ($R=2$), 
where wrapping issues appear first.
This example reproduces,
in a fairly simplified version, one of the situations one encounters in the 
more complicated case of $\mathcal{N}=4$ SYM. 

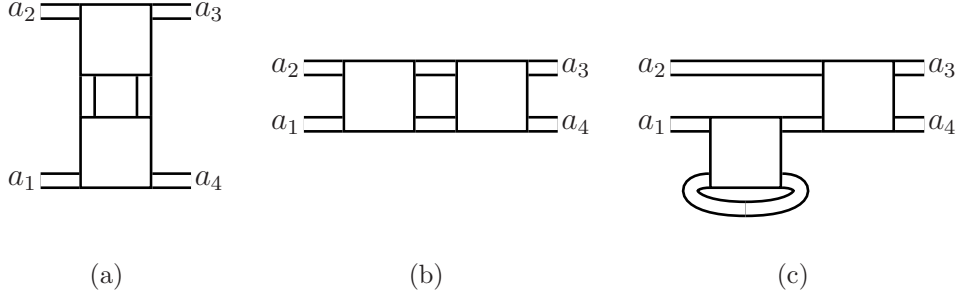
\begin{figure}
\begin{center}
\subfigure[]{\label{diagtype1}%
\begin{pspicture}(-1.5,-2)(13,11)%\showgrid
%\psframe(-1.5,-2)(13,11)
\setlength{\xa}{4.5\unit}
\addtolength{\xa}{-0.5\doublesep}
\addtolength{\xa}{-\linew}
\setlength{\xb}{7.5\unit}
\addtolength{\xb}{0.5\doublesep}
\addtolength{\xb}{\linew}
\setlength{\ya}{9\unit}
\addtolength{\ya}{0.5\doublesep}
\addtolength{\ya}{\linew}
\setlength{\yb}{6\unit}
\addtolength{\yb}{-0.5\doublesep}
\addtolength{\yb}{-\linew}
\setlength{\yc}{3\unit}
\addtolength{\yc}{0.5\doublesep}
\addtolength{\yc}{\linew}
\setlength{\yd}{0\unit}
\addtolength{\yd}{-0.5\doublesep}
\addtolength{\yd}{-\linew}
\psframe(\xa,\ya)(\xb,\yb)%\rput(0,-10.5){$\textstyle V_{2R}$}
\psframe(\xa,\yd)(\xb,\yc)
\psset{doubleline=true}
\psline(2,9)(\xa,9)
\psline(2,0)(\xa,0)
\psline(\xb,0)(10,0)
\psline(\xb,9)(10,9)
\psline(4.5,\yc)(4.5,\yb)
\psline(7.5,\yc)(7.5,\yb)
\rput[r](1.8,0){$a_1$}
\rput[r](1.8,9){$a_2$}
\rput[l](10.2,9){$a_3$}
\rput[l](10.2,0){$a_4$}
\end{pspicture}}%
\subfigure[]{\label{diagtype2}%
\begin{pspicture}(-4,-2)(15.5,11)%\showgrid
%\psframe(-4,-2)(15.5,11)
\setlength{\xa}{1.5\unit}
\addtolength{\xa}{-0.5\doublesep}
\addtolength{\xa}{-\linew}
\setlength{\xb}{4.5\unit}
\addtolength{\xb}{0.5\doublesep}
\addtolength{\xb}{\linew}
\setlength{\xc}{7.5\unit}
\addtolength{\xc}{-0.5\doublesep}
\addtolength{\xc}{-\linew}
\setlength{\xd}{10.5\unit}
\addtolength{\xd}{0.5\doublesep}
\addtolength{\xd}{\linew}
\setlength{\ya}{6\unit}
\addtolength{\ya}{0.5\doublesep}
\addtolength{\ya}{\linew}
\setlength{\yb}{3\unit}
\addtolength{\yb}{-0.5\doublesep}
\addtolength{\yb}{-\linew}
\psframe(\xa,\ya)(\xb,\yb)%\rput(0,-10.5){$\textstyle V_{2R}$}
\psframe(\xc,\ya)(\xd,\yb)
\psset{doubleline=true}
\psline(-1,6)(\xa,6)
\psline(-1,3)(\xa,3)
\psline(\xd,3)(12.5,3)
\psline(\xd,6)(12.5,6)
\psline(\xb,6)(\xc,6)
\psline(\xb,3)(\xc,3)
\rput[r](-1.2,3){$a_1$}
\rput[r](-1.2,6){$a_2$}
\rput[l](12.7,6){$a_3$}
\rput[l](12.7,3){$a_4$}
\end{pspicture}}%
\subfigure[]{\label{diagtype3}%
\begin{pspicture}(-4,-2)(15.5,11)%\showgrid
%\psframe(-4,-2)(15.5,11)
\setlength{\xa}{1.5\unit}
\addtolength{\xa}{-0.5\doublesep}
\addtolength{\xa}{-\linew}
\setlength{\xb}{4.5\unit}
\addtolength{\xb}{0.5\doublesep}
\addtolength{\xb}{\linew}
\setlength{\xc}{7.5\unit}
\addtolength{\xc}{-0.5\doublesep}
\addtolength{\xc}{-\linew}
\setlength{\xd}{10.5\unit}
\addtolength{\xd}{0.5\doublesep}
\addtolength{\xd}{\linew}
\setlength{\ya}{6\unit}
\addtolength{\ya}{0.5\doublesep}
\addtolength{\ya}{\linew}
\setlength{\yb}{3\unit}
\addtolength{\yb}{-0.5\doublesep}
\addtolength{\yb}{-\linew}
\psset{origin={0,3}}
\psframe(\xa,\ya)(\xb,\yb)%\rput(0,-10.5){$\textstyle V_{2R}$}
\psset{origin={0,0}}
\psframe(\xc,\ya)(\xd,\yb)
\psset{doubleline=true}
\psline(-1,6)(\xc,6)
%\psline(-1,6)(\xa,6)
\psbezier(\xa,0)(-0.5,0)(-0.5,-1.5)(3,-1.5)
\psbezier(\xb,0)(6.5,0)(6.5,-1.5)(3,-1.5)
\psline(-1,3)(\xa,3)
\psline(\xd,3)(12.5,3)
\psline(\xd,6)(12.5,6)
%\psline(\xb,6)(\xc,6)
\psline(\xb,3)(\xc,3)
\rput[r](-1.2,3){$a_1$}
\rput[r](-1.2,6){$a_2$}
\rput[l](12.7,6){$a_3$}
\rput[l](12.7,3){$a_4$}
\end{pspicture}}
\caption{The three possible types of connected ($C=1$) Feynman 
diagrams occurring at the perturbative order $g^4$ in $V_4$.} 
\label{diagtypes}
\end{center}
\end{figure}  

In a theory with vertices only of type \eqref{1} at order $g^4$, there are 
only three possible types of connected ($C=1$) Feynman diagrams contributing to
$V_4$.
We have depicted them in Fig.\ \ref{diagtypes}. 
Combining the six terms of the two $4$-vertices \eqref{2}
together, and considering all possible permutations of the external
legs, one gets $(4!)^2$ diagrams for the $t$- and the $u$-channel  
(Fig.\ \ref{diagtype1}) and for the $s$-channel (Fig.\
\ref{diagtype2}). The self energy contribution Fig.\ \ref{diagtype3}
leads to $4(4!)^2$ diagrams. All terms can be grouped together into $17$ 
different color structures that enter the full vertex obeying
crossing symmetry.
The result reads 
\begin{equation}
\begin{aligned} 
\label{4a}
V_{4,\text{sym.}}^{a_1 a_2 a_3
  a_4}&{}=\smash{\frac{2g^4}{(3!)^2}}\big[N\big\{(\alpha_t+\alpha_s+8\gamma')\big[\big(a_4\,a_1\,a_2\,a_3\big)+\big(a_4\,a_3\,a_2\,a_1\big)\big]\\
&\hphantom{={}\frac{2g^4}{(3!)^2}\big[N\big\{}+(\alpha_t+\alpha_u+8\gamma')\big[\big(a_4\,a_1\,a_3\,a_2\big)+\big(a_4\,a_2\,a_3\,a_1\big)\big]\\ 
&\hphantom{={}\frac{2g^4}{(3!)^2}\big[N\big\{}+(\alpha_u+\alpha_s+8\gamma')\big[\big(a_4\,a_3\,a_1\,a_2\big)+\big(a_4\,a_2\,a_1\,a_3\big)\big]\big\} \\ 
&\hphantom{={}\frac{2g^4}{(3!)^2}\big[{}}+(\alpha_t+\alpha_u+\alpha_s+3\gamma')\big\{
\big(a_1\,a_2\,a_3\big)\big(a_4\big)+\big(a_1\,a_3\,a_2\big)\big(a_4\big)\\
&\hphantom{={}\frac{2g^4}{(3!)^2}\big[+(\alpha_t+\alpha_u+\alpha_s+3\gamma')\big\{}+\big(a_1\,a_2\,a_4\big)\big(a_3\big)+\big(a_2\,a_1\,a_4\big)\big(a_3\big)\\
&\hphantom{={}\frac{2g^4}{(3!)^2}\big[+(\alpha_t+\alpha_u+\alpha_s+3\gamma')\big\{}+\big(a_1\,a_3\,a_4\big)\big(a_2\big)+\big(a_3\,a_1\,a_4\big)\big(a_2\big)\\
&\hphantom{={}\frac{2g^4}{(3!)^2}\big[{}+(\alpha_t+\alpha_u+\alpha_s+3\gamma')\big\{}+\big(a_2\,a_3\,a_4\big)\big(a_1\big)+\big(a_3\,a_2\,a_4\big)\big(a_1\big)\big\}\\
&\hphantom{={}\frac{2g^4}{(3!)^2}\big[}+(4\alpha_t+\alpha_u+\alpha_s)\big(a_1\,a_4\big)\big(a_2\,a_3\big)\\
&\hphantom{={}\frac{2g^4}{(3!)^2}\big[}+(\alpha_t+4\alpha_u+\alpha_s)\big(a_1\,a_3\big)\big(a_2\,a_4\big)\\
&\hphantom{={}\frac{2g^4}{(3!)^2}\big[}+(\alpha_t+\alpha_u+4\alpha_s)\big(a_1\,a_2\big)\big(a_3\,a_4\big)\big]
\col
\end{aligned}
\end{equation}
where $\alpha_t$, $\alpha_u$, $\alpha_s$, $\gamma'$ describe the
spacetime factors, including integrals over the positions of the two
vertices and depending on the coordinates $x_1$, $x_2$ and $y_1$,
$y_2$ of the ingoing and outgoing fields, respectively.
To obtain \eqref{4a} we have repeatedly used the $U(N)$ fusion and
fission rules \eqref{ffrulesshortnot} and the trace of the identity
$\big(\unitmatrix\big)=N$.

To match our definition of the Green function, when used as a building
block of  the $2$-point function, 
which is given in Subsection \ref{2ptexpansion} (see
also Fig.\ \ref{2ptfunctiondecomp}), we divide out the cyclic
permutations acting separately on the incoming and outgoing legs. In
this case these are simply the exchanges $a_1\leftrightarrow a_2$ and 
$a_3\leftrightarrow a_4$, under which the $17$ color terms in 
\eqref{4a} boil down to $6$ distinct structures.
We obtain the building block 
\begin{equation}
\begin{aligned} 
\label{4}
V_{4}^{a_1 a_2 a_3
  a_4}&{}=\smash{\frac{2g^4}{(3!)^2}}\big[N\big\{(\alpha+\beta+8\gamma)\big(a_4\,a_3\,a_2\,a_1\big)
%&\hphantom{={}\frac{2g^4}{(3!)^2}\big[N\big\{}
+(\alpha+4\gamma)\big(a_4\,a_1\,a_3\,a_2\big)\big\} \\ 
&\hphantom{={}\frac{2g^4}{(3!)^2}\big[{}}+(2\alpha+\beta+3\gamma)\big\{
\big(a_1\,a_2\,a_3\big)\big(a_4\big)+\big(a_1\,a_3\,a_4\big)\big(a_2\big)\big\}
\\
&\hphantom{={}\frac{2g^4}{(3!)^2}\big[}+\frac{1}{2}\big\{(5\alpha+\beta)\big(a_1\,a_4\big)\big(a_2\,a_3\big)+(\alpha+2\beta)\big(a_1\,a_2\big)\big(a_3\,a_4\big)\big\}\big]
\pnt
\end{aligned}
\end{equation}
Furthermore, we have identified $x_1=x_2=x$ and $y_1=y_2=y$ such that 
$\alpha_t,\alpha_u\to\alpha$, $\alpha_s\to\beta$ and $\gamma'\to\gamma$.
The coefficients $\alpha$, $\beta$, $\gamma$ contain the corresponding $3$-loop
integrals.
With the coordinate expression for the propagator
\begin{equation}
I_{xy}=\frac{\Gamma(\frac{d}{2}-1)}{4\pi^\frac{d}{2}}\frac{1}{((x-y)^2-i\epsilon)^{\frac{d}{2}-1}} 
\end{equation}
they read in $d=4-2\varepsilon$ dimensions
\begin{equation}
%\begin{aligned}
\alpha=\int\de^du\de^dv\,I_{xu}I_{xv}I_{uv}^2I_{uy}I_{vy}\col\qquad
\beta=\int\de^du\de^dv\,I_{xu}^2I_{uv}^2I_{vy}^2\col\qquad
\gamma=0\pnt
%\end{aligned}
\end{equation}
In the last equality the vanishing of all tadpole-type graphs in dimensional
regularization has been used. Moreover, they would not in any case
affect the analysis of planar wrapping diagrams, since it is
straightforwardly seen that they never lead to planar wrapping contributions. 
Explicit results for $\alpha$ and $\beta$ can be obtained by the 
Gegenbauer $x$-space technique (see (2.20) in \cite{Chetyrkin:1980pr}), and for
$\beta$ more easily by the methods based on uniqueness \cite{Kazakov:1986mu}.

The $2$-point function of the operator 
$\O{}{2}=\tr(\phi^2)$ is obtained from \eqref{4} by 
contracting the indices $a_1$, $a_2$ and $a_3$, $a_4$ with tensors 
proportional to $\delta^{a_1}_{a_2}$ and
$\delta^{a_3}_{a_4}$, respectively. This is obvious after rewriting 
$\tr(\phi^2)$ as $\phi_a \phi_b \tr(T^aT^b)$ and using \eqref{Tcommandtr}.
The color part reduces to either $N^4$ or $N^2$ if the 
corresponding diagram is of genus $H=0$ or $H=1$, respectively. This
is in accord with the genus expansion of the $2$-point function in 
Subsection \ref{2ptexpansion}. 

The planar contributions to the $2$-point function
are identified as the ones that contain the color structures 
$\big(a_4\,a_3\,a_2\,a_1\big)$ and $\big(a_1\,a_2\big)\big(a_3\,a_4\big)$.
By drawing the diagrams, one can check that the structure with a single 
trace is associated to the non-wrapping diagrams, while the 
double trace contribution takes care of the planar wrapping diagrams. This 
coincides with the generic results \eqref{planarV} and \eqref{planarVwrap}. 

The identification of planar wrapping diagrams is simple in this special case, 
where the number of diagrams is small enough to draw all of them.
We can therefore explicitly check our procedure 
based on spectator insertions as introduced in Subsection
\ref{spectatorfields}, adding two pairs of spectator fields such that the
modified $2$-point function becomes
\begin{equation}\label{7}
\begin{aligned}
\big(\rnode{1}{a_1}\,
\rnode{2}{\psi}\,\rnode{3}{a_2}\,\rnode{4}{\psi}\big)
\,V_4^{a_1 a_2 a_3 a_4}\, 
\big(\rnode{5}{\hat\psi}\,\rnode{6}{a_3}\,\rnode{7}{\hat\psi}\,
\rnode{8}{a_4}\big)+\text{cycl.\ perm.\ $\O{e}{2}$}
+\text{cycl.\ perm.\ $\hat\psi$}\pnt
\ncbar[linewidth=0.5pt,nodesep=1pt,angle=-90,arm=0.15cm]{4}{5}
\ncbar[linewidth=0.5pt,nodesep=1pt,angle=-90,arm=0.375cm]{2}{7}
\settodepth{\armlen}{$a_{R+1}$}
\addtolength{\armlen}{0.375cm}
\addtolength{\armlen}{1.5pt}
\rule[-\armlen]{0pt}{\armlen}
\end{aligned}
\end{equation}
Here, cyclic permutations within $\O{e}{2}$ corresponds to the $4$ possible 
combinations obtained by the changes $a_1\leftrightarrow a_2$,
$a_3\leftrightarrow a_4$ within the two operators $\O{e}{2}$.

The terms in \eqref{4} lead to the distinct spectator structures
\begin{equation}\label{8}
\begin{aligned}
%\settodepth{\armlen}{$a_{R+1}$}
%\addtolength{\armlen}{0.375cm}
%\addtolength{\armlen}{1.5pt}
%\renewcommand{\arraystretch}{2.5}
%\renewcommand{\extrarowheight}{-3pt}
%\renewcommand{\tabcolsep}{0pt}
%\setlength{\abovedisplayskip}{0pt plus 0pt minus 0pt}
%\setlength{\belowdisplayskip}{0pt plus 0pt minus 0pt}
%\begin{tabular}{@{\hspace{0pt}}r@{\hspace{0pt}}@{\hspace{0pt}}l@{\hspace{0pt}}}
\big(a_4\,a_3\,a_2\,a_1\big)&\to\big(\rnode{1}{\psi}\big)\big(\rnode{2}{\psi}\,\rnode{4}{\hat\psi}\big)\big(\rnode{3}{\hat\psi}\big)
\ncbar[linewidth=0.5pt,nodesep=1pt,angle=-90,arm=0.15cm]{2}{4}
\ncbar[linewidth=0.5pt,nodesep=1pt,angle=-90,arm=0.375cm]{1}{3}
+
\big(\rnode{1}{\psi}\big)\big(\rnode{2}{\psi}\,\rnode{4}{\hat\psi}\big)\big(\rnode{3}{\hat\psi}\big)
\col
\ncbar[linewidth=0.5pt,nodesep=1pt,angle=-90,arm=0.15cm]{2}{3}
\ncbar[linewidth=0.5pt,nodesep=1pt,angle=-90,arm=0.375cm]{1}{4}
\settodepth{\armlen}{$a_{R+1}$}
\addtolength{\armlen}{0.375cm}
\addtolength{\armlen}{1.5pt}
\rule[-\armlen]{0pt}{\armlen} 
\\
\big(a_4\,a_1\,a_3\,a_2\big)&\to\big(\rnode{1}{\psi}\,\rnode{2}{\hat\psi}\,\rnode{4}{\psi}\,\rnode{3}{\hat\psi}\big)
\ncbar[linewidth=0.5pt,nodesep=1pt,angle=-90,arm=0.15cm]{2}{4}
\ncbar[linewidth=0.5pt,nodesep=1pt,angle=-90,arm=0.375cm]{1}{3}
+
\big(\rnode{1}{\psi}\,\rnode{2}{\hat\psi}\,\rnode{4}{\psi}\,\rnode{3}{\hat\psi}\big)
\col
\ncbar[linewidth=0.5pt,nodesep=1pt,angle=-90,arm=0.15cm]{3}{4}
\ncbar[linewidth=0.5pt,nodesep=1pt,angle=-90,arm=0.15cm]{1}{2}
\settodepth{\armlen}{$a_{R+1}$}
\addtolength{\armlen}{0.375cm}
\addtolength{\armlen}{1.5pt}
\rule[-\armlen]{0pt}{\armlen} 
\\
\big(a_1\,a_2\,a_3\big)\big(a_4\big)&\to\big(\rnode{1}{\psi}\big)\big(\rnode{4}{\psi}\,\rnode{3}{\hat\psi}\,\rnode{2}{\hat\psi}\big)
\ncbar[linewidth=0.5pt,nodesep=1pt,angle=-90,arm=0.15cm]{2}{4}
\ncbar[linewidth=0.5pt,nodesep=1pt,angle=-90,arm=0.375cm]{1}{3}
+
\big(\rnode{1}{\psi}\big)\big(\rnode{4}{\psi}\,\rnode{3}{\hat\psi}\,\rnode{2}{\hat\psi}\big)
\col
\ncbar[linewidth=0.5pt,nodesep=1pt,angle=-90,arm=0.15cm]{3}{4}
\ncbar[linewidth=0.5pt,nodesep=1pt,angle=-90,arm=0.375cm]{1}{2}
\settodepth{\armlen}{$a_{R+1}$}
\addtolength{\armlen}{0.375cm}
\addtolength{\armlen}{1.5pt}
\rule[-\armlen]{0pt}{\armlen} 
\\
\big(a_1\,a_3\,a_4\big)\big(a_2\big)&\to\big(\rnode{1}{\psi}\,\rnode{2}{\psi}\,\rnode{4}{\hat\psi}\big)\big(\rnode{3}{\hat\psi}\big)
\ncbar[linewidth=0.5pt,nodesep=1pt,angle=-90,arm=0.15cm]{2}{4}
\ncbar[linewidth=0.5pt,nodesep=1pt,angle=-90,arm=0.375cm]{1}{3}
+
\big(\rnode{1}{\psi}\,\rnode{2}{\psi}\,\rnode{4}{\hat\psi}\big)\big(\rnode{3}{\hat\psi}\big)
\col
\ncbar[linewidth=0.5pt,nodesep=1pt,angle=-90,arm=0.15cm]{2}{3}
\ncbar[linewidth=0.5pt,nodesep=1pt,angle=-90,arm=0.375cm]{1}{4}
\settodepth{\armlen}{$a_{R+1}$}
\addtolength{\armlen}{0.375cm}
\addtolength{\armlen}{1.5pt}
\rule[-\armlen]{0pt}{\armlen} 
\\
\big(a_1\,a_4\big)\big(a_2\,a_3\big)&\to\big(\rnode{1}{\psi}\,\rnode{2}{\hat\psi}\big)\big(\rnode{4}{\psi}\,\rnode{3}{\hat\psi}\big)
\ncbar[linewidth=0.5pt,nodesep=1pt,angle=-90,arm=0.15cm]{2}{4}
\ncbar[linewidth=0.5pt,nodesep=1pt,angle=-90,arm=0.375cm]{1}{3}
+
\big(\rnode{1}{\psi}\,\rnode{2}{\hat\psi}\big)\big(\rnode{4}{\psi}\,\rnode{3}{\hat\psi}\big)
\col
\ncbar[linewidth=0.5pt,nodesep=1pt,angle=-90,arm=0.15cm]{3}{4}
\ncbar[linewidth=0.5pt,nodesep=1pt,angle=-90,arm=0.15cm]{1}{2}
\settodepth{\armlen}{$a_{R+1}$}
\addtolength{\armlen}{0.375cm}
\addtolength{\armlen}{1.5pt}
\rule[-\armlen]{0pt}{\armlen} 
\\
\big(a_1\,a_2\big)\big(a_3\,a_4\big)&\to2\,\big(\rnode{1}{\psi}\big)\big(\rnode{3}{\hat\psi}\big)\big(\rnode{2}{\psi}\big)\big(\rnode{4}{\hat\psi}\big)
\ncbar[linewidth=0.5pt,nodesep=1pt,angle=-90,arm=0.15cm]{2}{4}
\ncbar[linewidth=0.5pt,nodesep=1pt,angle=-90,arm=0.15cm]{1}{3}
%+
%\big(\rnode{1}{\psi}\big)\big(\rnode{3}{\hat\psi}\big)\big(\rnode{2}{\psi}\big)\big(\rnode{4}{\hat\psi}\big)
\pnt
%\ncbar[linewidth=0.5pt,nodesep=1pt,angle=-90,arm=0.15cm]{2}{3}
%\ncbar[linewidth=0.5pt,nodesep=1pt,angle=-90,arm=0.375cm]{1}{4}
\settodepth{\armlen}{$a_{R+1}$}
\addtolength{\armlen}{0.375cm}
\addtolength{\armlen}{1.5pt}
\rule[-\armlen]{0pt}{\armlen} 
%\rule[-\armlen]{0pt}{\armlen}
%&(4123) \longrightarrow (\psi_2 \psi_2)(\psi_1)(\psi_1),  \\
%&(4132) \longrightarrow (\psi_2 \psi_1 \psi_1 \psi_2),  \\
%&(4231) \longrightarrow (\psi_2 \psi_2 \psi_1 \psi_1),  \\
%&(4321) \longrightarrow (\psi_1 \psi_1)(\psi_2)(\psi_2),  \\
%&(4312) \longrightarrow (\psi_1 \psi_2)(\psi_1)(\psi_2),  \\
%&(4213) \longrightarrow (\psi_2 \psi_1)(\psi_1)(\psi_2),  \\
%&(412)(3) \longrightarrow (\psi_1 \psi_2 \psi_2)(\psi_1),  \\
%&(421)(3) \longrightarrow (\psi_1 \psi_1 \psi_2)(\psi_2),  \\
%&(413)(2) \longrightarrow (\psi_2 \psi_1 \psi_2)(\psi_1),  \\ 
%&(431)(2) \longrightarrow (\psi_2 \psi_1 \psi_1)(\psi_2),  \\
%&(234)(1) \longrightarrow (\psi_1 \psi_2 \psi_2)(\psi_1),  \\
%&(324)(1) \longrightarrow (\psi_1 \psi_2 \psi_1)(\psi_2),  \\
%&(123)(4) \longrightarrow (\psi_2 \psi_1 \psi_2)(\psi_1),  \\
%&(132)(4) \longrightarrow (\psi_1 \psi_2 \psi_1)(\psi_2),  \\
%&(14)(23) \longrightarrow (\psi_2 \psi_2)(\psi_1 \psi_1),  \\
%&(12)(34) \longrightarrow (\psi_1)(\psi_2)(\psi_2)(\psi_1),  \\
%&(13)(24) \longrightarrow (\psi_2 \psi_1)(\psi_2 \psi_1). 
\end{aligned}
\end{equation}
As can be easily seen, 
the only case in which the traces completely 
factorize into four single pieces is the wrapping case,
in accord with our general arguments in Subsection \ref{topencoding}. 

The toy example we presented up to here is interesting, but of course 
it does not include
the full $\mathcal{N}=4$ complexity. In order 
to get closer to that case, 
in a first step we add flavor degrees of freedom to the interaction Lagrangian
\eqref{1} in a simple way, such that the number of terms in the
symmetrized interaction vertex \eqref{2} is not enlarged.
In a second step, we then consider the $4$-scalar
commutator interaction of the full $\mathcal{N}=4$ theory, which symmetrized 
interaction vertex contains several terms with different flavor dependence.

\subsection{$4$-vertex with cyclic symmetric flavor flux}
\label{firstrefinement}
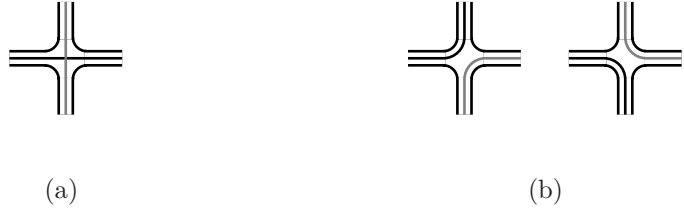
\begin{figure}
\begin{center}
\newlength{\flinearc}
\setlength{\flinearc}{\linearc}
\addtolength{\flinearc}{0.5\dlinewidth}
\subfigure[]{\label{flavorcrossing}%
\begin{pspicture}(-4,-4)(4,4)%\showgrid
%\psframe(-4,-4)(4,4)
\fourvertex{0}{0}{0}
\psset{doubleline=true}
\psline(-3,0)(-1,0)
\psline(0,-1)(0,-3)
\psline(1,0)(3,0)
\psline(0,1)(0,3)
\psset{doubleline=false}
\psline(-3,0)(0,0)
\psline(0,0)(3,0)
\psline[linecolor=gray](0,-3)(0,0)
\psline[linecolor=gray](0,0)(0,3)
\end{pspicture}}%
\qquad\qquad\qquad\qquad
\subfigure[]{\label{flavorrepulsing}%
\begin{pspicture}(-4,-4)(4,4)%\showgrid
%\psframe(-4,-4)(4,4)
\fourvertex{0}{0}{0}
\psset{doubleline=true}
\psline(-3,0)(-1,0)
\psline(0,-1)(0,-3)
\psline(1,0)(3,0)
\psline(0,1)(0,3)
\psset{doubleline=false}
\psline[linearc=\flinearc](-3,0)(0,0)(0,3)
\psline[linearc=\flinearc,linecolor=gray](0,-3)(0,0)(3,0)
\end{pspicture}
\begin{pspicture}(-4,-4)(4,4)%\showgrid
%\psframe(-4,-4)(4,4)
\fourvertex{0}{0}{0}
\psset{doubleline=true}
\psline(-3,0)(-1,0)
\psline(0,-1)(0,-3)
\psline(1,0)(3,0)
\psline(0,1)(0,3)
\psset{doubleline=false}
\psline[linearc=\flinearc](-3,0)(0,0)(0,-3)
\psline[linearc=\flinearc,linecolor=gray](0,3)(0,0)(3,0)
\end{pspicture}
}
\caption{The flow of a flavor line through a vertex of the types 
$\tr(\phi_m\phi_n\phi_m\phi_n)$ \subref{flavorcrossing} and
$\tr(\phi_m\phi_m\phi_n\phi_n)$ \subref{flavorrepulsing}.
In the second case, the two permutations have to be taken
  into account in the symmetrized vertex.}
\label{flavorflux}
\end{center}
\end{figure}

In our first refinement we take as interaction term 
\begin{equation}
\label{9}
S_\text{int}[\phi]=\frac{g^2}{4!}\tr(\phi_m\phi_n\phi_m\phi_n)\col
\end{equation} 
where we have introduced flavor indices $m,n=1,...,N_\text{f}$. 
The two flavor lines that enter and leave the vertex at its four legs cross each other as depicted in Fig.\ \ref{flavorcrossing}.

Since the structure with crossing flavor lines is invariant under cyclic
permutations, the symmetrized vertex consists of six inequivalent 
terms as in \eqref{2}, but with an additional simple factor for the
flavor dependence. One gets
\begin{equation}
\begin{aligned}
\label{10}
V^{a_1 a_2 a_3 a_4}_{m_1 m_2 m_3 m_4}&=\smash{\frac{g^2}{3!}}\big[\big(a_1\,a_2\,a_3\,a_4\big)\delta^{m_1}_{m_3}\delta^{m_2}_{m_4}+\big(a_1\,a_2\,a_4\,a_3\big)\delta^{m_1}_{m_4}  \delta^{m_2}_{m_3}\\
&\phantom{={}4g^2\big[}+\big(a_1\,a_3\,a_2\,a_4\big)\delta^{m_1}_{m_2}  
\delta^{m_3}_{m_4}+\big(a_2\,a_1\,a_3\,a_4\big)\delta^{m_1}_{m_4}\delta^{m_2}_{m_3}\\
&\phantom{={}4g^2\big[}
+\big(a_1\,a_4\,a_2\,a_3\big)\delta^{m_1}_{m_2}\delta^{m_3}_{m_4}+\big(a_1\,a_4\,a_3\,a_2\big)\delta^{m_1}_{m_3}\delta^{m_2}_{m_4}\big]\col 
\end{aligned}
\end{equation}
where $m_r$ is the corresponding flavor index carried by the leg with color
index $a_r$, arranged as in Fig.\ \ref{diagtypes}.

The genus expansion is an expansion in powers of $\frac{1}{N}$, the inverse
of the number of colors. Adding flavor degrees of freedom does not influence
this expansion, it simply extends the result of a single field by
multiplying each term by a flavor dependent factor.
We consider the $2$-point function of the operator
$\O{}{2}=\tr(\phi_m\phi_m)$ which now include a summation over all flavors.
The planar wrapping diagrams turn out to be given by
\begin{equation}
\label{11}
\frac{g^4}{(3!)^2}\big(a_1\,a_2\big)\big(a_3\,a_4\big)\big[(\alpha N_\text{f}\delta^{m_1}_{m_4}\delta^{m_2}_{m_3}+2\beta\delta^{m_1}_{m_3}\delta^{m_2}_{m_4}\big]\pnt 
\end{equation}
This result can be brought into the symmetrized form
\eqref{planarVwrapflavor}, where the single coefficient is
proportional to $\alpha N_\text{f}+2\beta$.

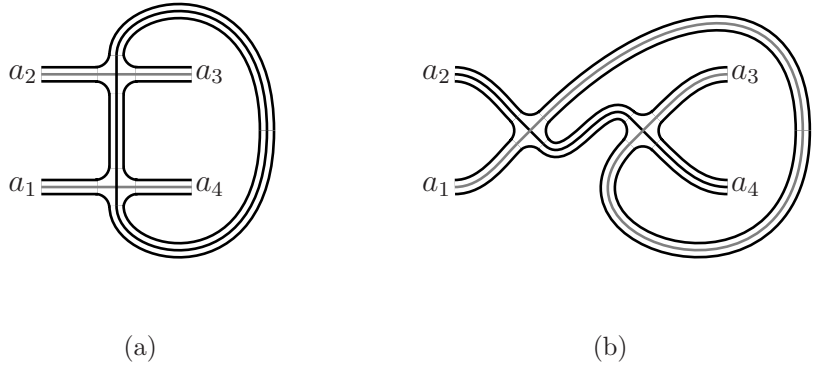
\begin{figure}
\begin{center}
\subfigure[]{\label{Nfloop}%
\begin{pspicture}(-5,-4)(20,13)%\showgrid
%\psframe(-5,-4)(20,13)
\fourvertex{6}{7.5}{0}
\fourvertex{6}{1.5}{0}
\setlength{\xa}{4.5\unit}
\addtolength{\xa}{-0.5\doublesep}
\addtolength{\xa}{-\linew}
\setlength{\xb}{7.5\unit}
\addtolength{\xb}{0.5\doublesep}
\addtolength{\xb}{\linew}
\setlength{\ya}{6\unit}
\addtolength{\ya}{0.5\doublesep}
\addtolength{\ya}{\linew}
\setlength{\yb}{6\unit}
\addtolength{\yb}{-0.5\doublesep}
\addtolength{\yb}{-\linew}
\setlength{\yc}{3\unit}
\addtolength{\yc}{0.5\doublesep}
\addtolength{\yc}{\linew}
\setlength{\yd}{3\unit}
\addtolength{\yd}{-0.5\doublesep}
\addtolength{\yd}{-\linew}
\psset{doubleline=true}
\psline(2,7.5)(5,7.5)
\psline(2,1.5)(5,1.5)
\psline(7,1.5)(10,1.5)
\psline(7,7.5)(10,7.5)
\psbezier(6,8.5)(6,11.5)(14,13)(14,4.5)
\psbezier(6,0.5)(6,-2.5)(14,-4)(14,4.5)
\psline(6,2.5)(6,6.5)
\psset{linecolor=gray,doubleline=false}
\psline(2,7.5)(10,7.5)
\psline(2,1.5)(10,1.5)
\psset{linecolor=black,doubleline=false}
\psbezier(6,8.5)(6,11.5)(14,13)(14,4.5)
\psbezier(6,0.5)(6,-2.5)(14,-4)(14,4.5)
\psline(6,0.5)(6,8.5)
\rput[r](1.8,1.5){$a_1$}
\rput[r](1.8,7.5){$a_2$}
\rput[l](10.2,7.5){$a_3$}
\rput[l](10.2,1.5){$a_4$}
\end{pspicture}}%
\subfigure[]{\label{noNfloop}%
\begin{pspicture}(-5,-4)(20,13)%\showgrid
%\psframe(-5,-4)(20,13)
\fourvertex{3}{4.5}{45}
\fourvertex{9}{4.5}{45}
\setlength{\xa}{1.5\unit}
\addtolength{\xa}{-0.5\doublesep}
\addtolength{\xa}{-\linew}
\setlength{\xb}{4.5\unit}
\addtolength{\xb}{0.5\doublesep}
\addtolength{\xb}{\linew}
\setlength{\xc}{7.5\unit}
\addtolength{\xc}{-0.5\doublesep}
\addtolength{\xc}{-\linew}
\setlength{\xd}{10.5\unit}
\addtolength{\xd}{0.5\doublesep}
\addtolength{\xd}{\linew}
\setlength{\ya}{6\unit}
\addtolength{\ya}{0.5\doublesep}
\addtolength{\ya}{\linew}
\setlength{\yb}{3\unit}
\addtolength{\yb}{-0.5\doublesep}
\addtolength{\yb}{-\linew}
%\psframe(\xa,\ya)(\xb,\yb)%\rput(0,-10.5){$\textstyle V_{2R}$}
%\psframe(\xc,\ya)(\xd,\yb)
\psset{doubleline=true}
\psbezier(-1,7.5)(0.5,7.5)(1.5,6)(2.2929,5.2071)
\psbezier(-1,1.5)(0.5,1.5)(1.5,3)(2.2929,3.7929)
\psbezier(13.5,7.5)(12,7.5)(11,6.5)(9.7071,5.2071)
\psbezier(13.5,1.5)(12,1.5)(11,2.5)(9.7071,3.7929)
\psbezier(3.7071,3.7929)(5.12132,2.2929)(6.87868,6.62132)(8.2929,5.2071)
\psbezier(3.7071,5.2071)(10.5,12)(17.5,12)(17.5,4.5)
\psbezier(8.2929,3.7929)(3,-1.5)(17.5,-6)(17.5,4.5)
\psset{linecolor=black,doubleline=false}
\psbezier(-1,7.5)(0.5,7.5)(1.5,6)(2.2929,5.2071)
\psbezier(13.5,1.5)(12,1.5)(11,2.5)(9.7071,3.7929)
\psline(2.2929,5.2071)(3.7071,3.7929)
\psline(8.2929,5.2071)(9.7071,3.7929)
\psbezier(3.7071,3.7929)(5.12132,2.2929)(6.87868,6.62132)(8.2929,5.2071)
\psset{linecolor=gray,doubleline=false}
\psbezier(-1,1.5)(0.5,1.5)(1.5,3)(2.2929,3.7929)
\psbezier(13.5,7.5)(12,7.5)(11,6.5)(9.7071,5.2071)
\psbezier(3.7071,5.2071)(10.5,12)(17.5,12)(17.5,4.5)
\psbezier(8.2929,3.7929)(3,-1.5)(17.5,-6)(17.5,4.5)
\psline(2.2929,3.7929)(3.7071,5.2071)
\psline(8.2929,3.7929)(9.7071,5.2071)
\rput[r](-1.2,1.5){$a_1$}
\rput[r](-1.2,7.5){$a_2$}
\rput[l](13.7,7.5){$a_3$}
\rput[l](13.7,1.5){$a_4$}
\end{pspicture}}
\caption{Contributions to planar wrapping interactions in the presence of
  flavor. The structure in Fig.\ \ref{diagtype1} leads to
  contributions with a closed flavor loop as shown in 
\subref{Nfloop}. The structure in Fig.\ \ref{diagtype2} leads to  
contributions depending on $\delta^{m_1}_{m_3}\delta^{m_2}_{m_4}$
  as shown in \subref{noNfloop}.}
\end{center}
\end{figure}

In the corresponding ribbon graph Fig.\ \ref{planarVwrapdiag} the color
structure
$\big(a_1\,a_2\big)\big(a_3\,a_4\big)$ has a simple interpretation as 
the breaking of the single surrounding index line by the structure along the
wrapping path. Adding flavor lines to the corresponding diagrams, one 
finds an interpretation even for the flavor structure in \eqref{11}. 
A vertex with the flavor flow as given in Fig.\ \ref{flavorcrossing} yields
the diagram depicted in Fig.\ \ref{Nfloop}. 
A factor $N_\text{f}$ can be traced back to a closed flavor loop which is
present in diagrams of the type shown in Fig.\ \ref{diagtype1}.
The remaining flavor lines can be associated to the factor  
$\delta^{m_1}_{m_4}\delta^{m_2}_{m_3}$.
The other contributions are based on the type of diagrams  
shown in Fig.\ \ref{diagtype2}. They do not contain closed flavor
loops and can be associated to the flavor structure
$\delta^{m_1}_{m_3}\delta^{m_2}_{m_4}$, see Fig.\ \ref{noNfloop}. 

\subsection{$4$-vertex with commutators}
\label{secondrefinement}
Now we get more into contact with $\mathcal{N}=4$ SYM adopting the 
proper commutator interaction given by
\begin{equation}
\label{12}
S_\text{int}[\phi]=\frac{1}{2}\frac{g^2}{4!}\tr[\phi_m,\phi_n][\phi_m,\phi_n]= 
\frac{g^2}{4!}\tr(\phi_m\phi_n\phi_m\phi_n)-\frac{g^2}{4!}\tr(\phi_m\phi_m\phi_n\phi_n)\col
\end{equation} 
where we have introduced the factor $\frac{1}{2}$ for convenience.

The interaction in \eqref{12}
appears as the combination of the previous one \eqref{9} with another term 
in which the flavor lines `repulse' each other, as depicted in Fig.\
\ref{flavorrepulsing}. 

The vertex with repulsing flavor lines is not invariant under all
cyclic permutations, but only under those where the legs are shifted by
an even number. So one has to keep track 
of two contributions, differing by an odd number of shifts. 
The explicit form of the vertex turns out to be
\begin{equation}
\begin{aligned}
\label{13}
V^{a_1 a_2 a_3 a_4}_{m_1 m_2 m_3
  m_4}&=\smash{\frac{2g^2}{4!}}\big[\big(a_1\,a_2\,a_3\,a_4\big)(2\delta^{m_1}_{m_3}\delta^{m_2}_{m_4}-\delta^{m_1}_{m_2}\delta^{m_3}_{m_4}-\delta^{m_1}_{m_4}\delta^{m_2}_{m_3})
  \\ 
&\hphantom{={}\frac{2g^2}{4!}\big[}+\big(a_1\,a_2\,a_4\,a_3\big)(2\delta^{m_1}_{m_4}\delta^{m_2}_{m_3}-\delta^{m_1}_{m_3}\delta^{m_2}_{m_4}-\delta^{m_1}_{m_2}\delta^{m_3}_{m_4})
  \\
&\hphantom{={}\frac{2g^2}{4!}\big[}+\big(a_1\,a_3\,a_2\,a_4\big)(2\delta^{m_1}_{m_2}\delta^{m_3}_{m_4}-\delta^{m_1}_{m_3}\delta^{m_2}_{m_4}-\delta^{m_1}_{m_4}\delta^{m_2}_{m_3})
  \\ 
&\hphantom{={}\frac{2g^2}{4!}\big[}+\big(a_2\,a_1\,a_3\,a_4\big)(2\delta^{m_1}_{m_4}\delta^{m_2}_{m_3}-\delta^{m_1}_{m_3}\delta^{m_2}_{m_4}-\delta^{m_1}_{m_2}\delta^{m_3}_{m_4})
  \\
&\hphantom{={}\frac{2g^2}{4!}\big[}+\big(a_1\,a_4\,a_2\,a_3\big)(2\delta^{m_1}_{m_2}\delta^{m_3}_{m_4}-\delta^{m_1}_{m_3}\delta^{m_2}_{m_4}-\delta^{m_1}_{m_4}\delta^{m_2}_{m_3})
  \\ 
&\hphantom{={}\frac{2g^2}{4!}\big[}+\big(a_1\,a_4\,a_3\,a_2\big)(2\delta^{m_1}_{m_3}\delta^{m_2}_{m_4}-\delta^{m_1}_{m_2}\delta^{m_3}_{m_4}-\delta^{m_1}_{m_4}\delta^{m_2}_{m_3})\big]\pnt
\end{aligned}
\end{equation}
The identification of the planar wrapping diagrams proceeds as in Subsection
\ref{firstrefinement}. After a tedious but straightforward computation
of the flavor factor one ends up with the result
\begin{equation}
\label{14}
\frac{g^4}{2(3!)^2}\big(a_1\,a_2\big)\big(a_3\,a_4\big)\big[(\alpha-2\beta+\beta N_\text{f})\delta^{m_1}_{m_2}\delta^{m_3}_{m_4}+(\alpha+\beta)\delta^{m_1}_{m_3}\delta^{m_2}_{m_4}+2\alpha(N_\text{f}-2)\delta^{m_1}_{m_4}\delta^{m_2}_{m_3}\big]\pnt  
\end{equation}
Its flavor dependence can again be reconstructed by taking into account all the
possible flavor flows generated by combining the vertices in Fig.\
\ref{flavorflux}. This result can be brought into the symmetrized form
\eqref{planarVwrapflavor}, where the single coefficient is
proportional to $2\alpha N_\text{f}-3\alpha+\beta$.

\section{A candidate for $\psi$}
\label{candpsi}

In this Section we want to show how one can project out the planar
wrapping contributions from all contributions to the $2$-point
function that contain $\V{}{1}{2R}$, 
by choosing suitable matrices for the spectator pairs $\psi$, $\hat\psi$.
The planar wrapping contributions have the property that they are the
only contributions to $\V{}{1}{2R}$ that generate the spectator
structure \eqref{planarwrapspecdiag} where the trace of all spectator
fields is taken separately. If one could find a non traceless
matrix with the property that its positive powers $p$ are traceless up to a
sufficiently high order $p_0$, $2\le p\le p_0$, 
then one could replace all spectator fields by this matrix. The 
non-wrapping contributions to $\V{}{1}{2R}$ would then vanish
automatically when used to build the $2$-point function of these
modified operators $\O{e}{2R}$.

In order to find such a matrix, we parameterize $\psi$, $\hat\psi$ as 
$\psi=\hat\psi=\psi_0T^0+\psi_aT^a$, $a=1,...,N^2-1$ with respect to the
decomposition of $U(N)$ into a $U(1)$ part and an $SU(N)$ part. 
The coefficients $\psi_0$ and $\psi_a$ are complex. 
A non traceless $\psi$ requires that $\psi_0 \neq 0$. 
The condition that 
$\tr(\psi^2) = 0$ amounts to the `light-cone' condition
\begin{equation}
\label{15}
\psi_a\psi^a+\psi_0^2=0\col
\end{equation}
where we have used \eqref{Tcommandtr} and \eqref{TU1}. 
The vanishing of the traces that contain higher powers of
$\psi$ is greatly 
simplified by the recursive use of the relations found up to the step before.
For example, evaluation of $\tr(\psi^3)$ and use of  \eqref{15} gives
\begin{equation}
\label{16}
\psi_a\psi_b\psi_c\tr(T^aT^bT^c)-\frac{2}{\sqrt{N}}\psi_0^3=0\pnt
\end{equation}    
After a little thought, one realizes that the general form of this equation at
order $n$ is 
\begin{equation}
\label{17}
\psi_{a_1}\dots\psi_{a_n}\tr(T^{a_1}\dots T^{a_n})+\frac{c_n}{\sqrt{N}^{n-2}}\psi_0^n=0\col
\end{equation}
where the $c_n$ are determined by recursion as
\begin{equation}
\label{18}
c_n=1-\binom{n}{2}-\sum_{m=3}^{n-1}\binom{n}{m}c_m
\end{equation}                
for $n \geq 3$ (for $n=3$ the sum has to be considered empty). 
This recursion is solved by 
\begin{equation}
\label{19}
c_n={(- 1)}^n(n - 1)\col 
\end{equation}
as can be easily verified using properties of binomial sums. The constraint
on $\psi$ is given by \eqref{15} combined with
\begin{equation}
\label{20}
\psi_{a_1}\dots\psi_{a_n}\tr(T^{a_1}\dots T^{a_n})+\frac{(-1)^n}{\sqrt{N}^{n-2}}(n-1)\psi_0^n=0
\end{equation}
for $n=3,\dots,n_0$ up to a desired order, which coincides with the 
highest possible power of $\psi$ within a single trace, obviously
given by $n_0=2R$.

The choice $\psi_0=\frac{1}{\sqrt{N}}$ normalizes the trace
$\big(\psi\big)=1$,  such that the additional matrices within the
traces do not change the prefactor for the planar wrapping interactions. 
The conditions \eqref{15} and \eqref{20} then read  
\begin{equation}
\label{20a}
\psi_a\psi^a=-\frac{1}{N}\col\qquad\psi_{a_1}\dots\psi_{a_n}\tr(T^{a_1}\dots T^{a_n})+\frac{(-1)^n}{N^{n-1}}(n-1)=0\pnt
\end{equation}

The important question to answer is whether this 
set of conditions has a solution in general.
This involves determining the components of the tensors  
$k^{a_1\dots a_n}=\tr(T^{(a_1}\dots T^{a_n)})$
appearing in \eqref{20}, where the parentheses indicate complete 
symmetrization with unit weight (i.e. summing over all permutations, and 
dividing by $n!$). These objects are a family of symmetric invariant
tensors of $SU(N)$ \cite{deAzcarraga:1997ya}. Not all of them are
independent. For   
$n\geq N+1$ $k^{a_1\dots a_n}$ can be expressed in
terms of the lower order ones. 
Instead of working with these objects, we have tried to test whether it
is possible to find any matrix $M$ with a non-vanishing trace, such
that $M^2$, $M^3$ and $M^4$ have vanishing trace. We were able to 
find a diagonal matrix of rank $12$ with pairs of complex conjugated 
eigenvalues $\lambda_i$,$\bar\lambda_i$ of unit length given by
\begin{equation}
\lambda_{1,2}=-1\col\qquad\lambda_3=\frac{1+i\sqrt{3}}{2}\col\qquad\lambda_{4,5,6}=\frac{-1+i\sqrt{3}}{2}\pnt
\end{equation} 

\section{Conclusions}
In this paper we have studied the genus expansion of the $2$-point function
of single-trace operators containing $R$ elementary fields, and of one of
its building blocks which is the 
Green function with $2R$ external legs.
We have shown that the two genus expansions do not coincide. 
The contributions to the Green 
function of a fixed genus $h$ lead to genus $h-2\leq H\leq h$ contributions to 
the $2$-point function. We have discussed the two origins for this effect.
Firstly, a reduction of the genus is caused by the invariance of the
trace of the composite operators under cyclic permutations. 
Secondly, a reduction of the genus is caused by the fact that certain
crossings between field lines in some contributions to the Green
function can be resolved
when it is used for the construction of the $2$-point function.  
In these wrapping contributions some field lines occupy a special (wrapping)
path, present in case of the $2$-point function, thereby avoiding 
some crossings that would require adding a handle.
These wrapping contributions play a role whenever the range of an
interaction becomes larger
than the length of the object it acts on. In the context
of the $\AdS/\text{CFT}$ correspondence it is believed that
they can explain an observed mismatch between the energies of
classical strings and the anomalous dimensions eigenvalues of the
corresponding operators starting at three loops. 

To avoid the additional genus changes caused by the cyclic symmetry of
the local operators, we have 
introduced equivalence classes for the diagrams of the Green function. 
Diagrams that differ only by cyclic permutations within the 
ingoing and outgoing legs are identified, and for the construction of
the $2$-point function only one diagram with minimal genus $h$ is
taken from each equivalence class. 
As a result of the analysis we have found that
all genus $H$ contributions to the $2$-point function of composite 
single-trace operators are obtained from these genus $h=H$ and $h=H+1$
contributions to the Green function. 

In our opinion this observation should be of particular importance.
At a given sufficiently high order in the coupling constant, one 
can divide the contributions to the $2$-point function and to the Green
function into two classes. 
One class contains those contributions which are universal, i.e. they 
have to be used in the construction of the $2$-point function of
operators with a larger number of legs. The second class contains 
those contributions which are non universal. 
Its elements change whenever the
number of legs of the operators is changed.
However, our above given result shows that even these non universal
contributions are contained within a universal quantity, the Green
function. Thus, the non universality of the elements within the second
class can be understood as a non universality of the projection
operation used to obtain these elements.
A better understanding of the projection operation might therefore
provide a way to obtain the wrapping contributions without the need to 
implement a selection process on each diagram separately. 

As a first step into this direction, we have worked out a technique,
that allows one to identify the wrapping contributions in the planar
($H=0$) case, which is of particular interest in the context of
the $\AdS/\text{CFT}$ correspondence. The technique is based on 
adding pairs of spectator fields and their pairwise connections to the 
diagrams of the $2$-point function.
The planar wrapping contributions are then identified as the 
only contributions in which, after contracting all the other fields, 
the trace over each matrix-valued spectator field is taken separately. 
A particular choice for the 
components of the spectator fields should therefore 
project out the planar wrapping contributions automatically. We have 
determined 
the equations which such a matrix must fulfill and we have presented 
a simple example to demonstrate that the problem itself should have a 
solution. 

Clearly, our analysis is only a first step to a complete
understanding of the wrapping interactions. As a next step one
could try to restrict the freedom in our proposed general 
vertex that should describe wrapping interactions in the traceless $SU(2)$
subsector. It would be interesting to discuss whether such an
interaction term can be compatible with integrability.   

%%% Local Variables:
%%% mode: latex
%%% TeX-master: "wrap"
%%% End: 

\section*{Acknowledgements}
\addcontentsline{toc}{section}{Acknowledgements}
We would like to thank B.\ Eden, J.\ Plefka and E.\ Sokatchev for deep and 
enlightening discussions.  
The work of C.\ S.\ is supported by DFG (Deutsche Forschungsgemeinschaft)
within `Graduiertenkolleg 271' and within project ER\,301/1-4.
The work of A.\ T.\ is supported by DFG within the
`Schwerpunktprogramm Stringtheorie 1096'. 
\appendix
\section{Counting rules for Feynman diagrams}
\label{countingrules}
We use the quantities
\begin{equation}\label{diagquantities}
\begin{aligned}
K&=\text{order of the diagram $g^K$}\col\\
E&=\text{number of external lines}\col\\
P&=\text{number of propagators}\col\\
V^k_i&=\text{number of vertices of order $g^k$ with $i$ legs}\col\\
V&=\text{number of vertices}\col\\
L&=\text{number of loops}\col\\
C&=\text{number of connected pieces}\col
\end{aligned}
\end{equation}
that classify a generic Feynman diagram.
The following relations hold
\begin{equation}\label{diageq}
%\begin{aligned}
K=\sum_{i,k}kV^k_i\col\;\:
V=\sum_{i,k}V^k_i\col\;\:
E=\sum_{i,k}iV^k_i-2P\col\;\:
P=V+L-C\pnt 
%\end{aligned}
\end{equation}
The equation for $E$ can be cast into the form
\begin{equation}\label{E}
%\begin{aligned}
E=\sum_{i,k}(i-2)V^k_i-2(L-C)\pnt
%\end{aligned}
\end{equation}
It simplifies if the $V^k_i$ obey
\begin{equation}\label{specialvertex}
V^k_i=V^k_{k+2}\delta_{i,k+2}\col
\end{equation}
i.e. from the coupling constant $g$ of the three point vertex all coupling 
constants of the higher point vertices are uniquely determined by adding one 
power in $g$ for each additional leg.  
For example, in YM theories one only has the 
cubic and quartic vertices, and therefore respectively $V^1_3$ and 
$V^2_4$ are different from zero.
With \eqref{specialvertex} one finds that \eqref{diageq} simplifies to
\begin{equation}\label{diageqspec}
%\begin{aligned}
K=\sum_kkV^k_{k+2}\col\;\:
V=\sum_kV^k_{k+2}\col\;\:
E=K-2(L-C)\col\;\:
P=V+L-C\pnt
%\end{aligned}
\end{equation}
Consider now the special case of a $2$-point function 
$\big(\O{1}{R},V_{2R},\O{2}{R}\big)$ of two composite
operators $\O{e}{R}$, where all vertices in $V_{2R}$ are of the form 
\eqref{specialvertex}. One can regard the operators as two vertices that do not
contribute to the order in the coupling constant and that have $R$ legs, 
i.e. $V^0_R=2$. Furthermore, one has $E=0$ since all legs of
$\O{1}{R}$ are contracted with the external legs of $V_{2R}$.
From the equation for $E$ in \eqref{diageq} one then finds in this case
\begin{equation}\label{Kof2pointfunc}
K=2(L-C-R+2)\pnt
\end{equation}
One can now determine the minimum order $K$ for a planar wrapping diagram
to appear. For this purpose, one has to estimate the minimum number of $L$.
The simplest non-interacting case in Subsection \ref{noninteractingcase} 
with $K=0$, $C=1$ inserted into
\eqref{Kof2pointfunc} gives $L=R-1$. In this case one can add $R$ spectators
to the diagram without making it non-planar by crossing any other lines. 
Therefore, each of the $R-1$ loops has to be divided into at least $2$-loops 
by a line that crosses a spectator line. Furthermore, one has to add at least
one line that runs along the wrapping path, increasing the loop number by at
least one. The number of loops in a wrapping diagram thus fulfills $L\ge2R-1$.
Clearly $C=1$, and \eqref{Kof2pointfunc} then leads to a lower bound for $K$
that becomes
\begin{equation}\label{Kboundcritwrap}
K\ge2R\pnt
\end{equation}
The structure of a generic planar wrapping diagram is arbitrarily
complicated. In particular, the interaction part that fills the
wrapping path can itself be an arbitrary planar diagram, see Fig.\
\ref{planarVwrapdiag}.
However, up to a certain order $K$ in the coupling constant, all wrapping 
diagrams can be obtained from 
\emph{planar non-wrapping connected diagrams of maximum interaction length}
by \emph{adding a single line} along the wrapping path. 

For a given wrapping diagram one can use the cyclic symmetry to minimize the 
number of lines that run along the wrapping path. One only has to guarantee
that the order $K$ in the coupling constant is sufficiently low such that 
at least one spectator line can be added that only crosses a single line 
of the planar diagram. This means all the other spectator lines in this 
case are allowed to cross at most two other lines. If one of them crosses more
than two other lines, one automatically finds a diagram at the same order 
where none of the spectator lines crosses only a single line. This is due to
the fact that one could replace the third crossed line by a line that
generates a second crossing of the spectator line that previously crossed only
one other line. Since each of the lines that connect neighboured fields of the
operators in a given diagram effectively contributes with a factor $g^2$, the
order $K$ for which all wrapping diagrams can be obtained by adding a single
line to the planar non-wrapping connected diagrams of maximum interaction
length has to fulfill 
\begin{equation}\label{Kboundonelegwrap}
K\le2(2(R-1)+1)=2(2R-1)\pnt
\end{equation}
 
\section{Rules for $U(N)$ and $SU(N)$}
\label{algebrarules}
Let $\T{a}{i}{j}$, $a=a_0,\dots N^2-1$, $i,j=1\dots N$ be the $N\times
N$ representation matrices for the Lie algebras of $U(N)$ (with $a_0=0$) 
and $SU(N)$ (with $a_0=1$). They are chosen such that they fulfill
\begin{equation}\label{Tcommandtr}
\comm{T^a}{T^b}=if^{abc}T^c\col\qquad
\tr(T^aT^b)=\delta^{ab}\col
\end{equation}
where the trace $\tr$ is taken over the fundamental indices $i,j$
respectively.
In particular, the $U(1)$ generator is given by
\begin{equation}\label{TU1}
T^0=\frac{1}{\sqrt{N}}\unitmatrix\pnt
\end{equation}
The representation matrices then fulfill the relation
\begin{equation}\label{Tcompleterel}
\sum_{a=a_0}^{N^2-1}\T{a}{i}{j}\T{a}{k}{l}=\delta^i_l\delta^k_j-\frac{a_0}{N}\delta^i_j\delta^k_l\col
\end{equation}
where as above $a_0=0$ and $a_0=1$ for $U(N)$ and $SU(N)$ respectively.
With the help of the above relation it is straigthforward to obtain the 
fusion and fission rules for the traces
\begin{equation}\label{ffrules}
\begin{aligned}
\tr(T^aA)\tr(T^aB)&=\tr(AB)-\frac{a_0}{N}\tr(A)\tr(B)\col\\
\tr(T^aA\,T^aB)&=\tr(A)\tr(B)-\frac{a_0}{N}\tr(AB)\pnt
\end{aligned}
\end{equation}
In the main text we use an abbreviated notation where the traces are simply
represented by parentheses and indices are no
longer written as superscripts. The above rules then read
\begin{equation}\label{ffrulesshortnot}
\begin{aligned}
\big(a\,A\big)\big(a\,B\big)&=\big(A\,B\big)-\frac{a_0}{N}\big(A\big)\big(B\big)\col\\
\big(a\,A\,a\,B\big)&=\big(A\big)\big(B\big)-\frac{a_0}{N}\big(A\,B\big)\pnt
\end{aligned}
\end{equation} 

%%% Local Variables:
%%% mode: latex
%%% TeX-master: "wrap"
%%% End: 

\footnotesize
\bibliographystyle{utphys}
\addcontentsline{toc}{section}{Bibliography}
\bibliography{references_BMN,references_wrap}

\end{document}